\documentclass[prl,nobalancelastpage,twocolumn,superscriptaddress,longbibliography]{revtex4-2}

\usepackage{color,amsthm,amsmath,amsxtra,amsfonts,dsfont,graphicx,bm,amssymb,mathtools}
\usepackage[colorlinks=true,linkcolor=blue, citecolor=blue, urlcolor=blue, bookmarks]{hyperref}
\usepackage{centernot}
\usepackage[dvipsnames]{xcolor}
\usepackage{graphicx}
\usepackage{tikz}
\usepackage{braket}
\usepackage{multirow}
\usepackage{makecell}

\definecolor{oceanboatblue}{rgb}{0.0, 0.47, 0.75}

\newcommand{\be}{\begin{equation}}
	\newcommand{\ee}{\end{equation}}
\newcommand{\bea}{\begin{eqnarray}}
	\newcommand{\eea}{\end{eqnarray}}
\newcommand{\bse}{\begin{subequations}}
	\newcommand{\ese}{\end{subequations}}

\theoremstyle{plain}

\newcommand{\prlsection}[1]{{\em {#1}.---~}}

\usepackage{pifont}
\usepackage{tikz}
\usetikzlibrary{decorations.pathreplacing,calligraphy,decorations.markings}

\begin{document}
	\date{\today}
	
	\newcommand{\bbra}[1]{\<\< #1 \right|\right.}
	\newcommand{\kket}[1]{\left.\left| #1 \>\>}
	\newcommand{\bbrakket}[1]{\< \Braket{#1} \>}
	\newcommand{\pll}{\parallel}
	\newcommand{\nn}{\nonumber}
	\newcommand{\transp}{\text{transp.}}
	\newcommand{\nor}{z_{J,H}}
	
	\newcommand{\hL}{\hat{L}}
	\newcommand{\hR}{\hat{R}}
	\newcommand{\hQ}{\hat{Q}}

	\title{Entanglement dynamics and Page curves in random permutation circuits}

\begin{abstract}
The characterization of ensembles of many-qubit random states and their realization via quantum circuits are crucial tasks in quantum-information theory. In this work, we study the ensembles generated by quantum circuits that randomly permute the computational basis, thus acting classically on the corresponding states. We focus on the averaged entanglement and present two main results. First, we derive generically tight upper bounds on the entanglement that can be generated by applying permutation circuits to arbitrary initial states. We show that the late-time ``entanglement Page curves'' are bounded in terms of the initial state participation entropies and its overlap with the ``maximally antilocalized'' state. Generally speaking, this result states that the quantum correlations generated by classical circuits are bounded in terms of some quantum property of the initial state (namely, the degree to which it can be written as a superposition of classical states). Second, comparing the averaged R\'enyi-$2$ entropies generated by $(i)$ an infinitely deep random circuit of two-qubit gates and $(ii)$ global random permutations, we show that the two quantities are different for finite $N$ but the corresponding Page curves coincide in the thermodynamic limit. We also discuss how these conclusions are modified by additional random phases or considering circuits of $k$-local gates with $k\geq 3$. Our results are exact and highlight the implications of classical features on entanglement generation in many-body systems.
\end{abstract}

\author{D\'avid Sz\'asz-Schagrin}
\thanks{These authors contributed equally to this work.}
\affiliation{Dipartimento di Fisica e Astronomia, Universit\`a di Bologna and INFN, Sezione di Bologna, via Irnerio 46, I-40126 Bologna, Italy}

\author{Michele Mazzoni}
\thanks{These authors contributed equally to this work.}
\affiliation{Dipartimento di Fisica e Astronomia, Universit\`a di Bologna and INFN, Sezione di Bologna, via Irnerio 46, I-40126 Bologna, Italy}

\author{Bruno Bertini}
\affiliation{School of Physics and Astronomy, University of Birmingham, Edgbaston, Birmingham, B15 2TT, UK}
\author{Katja Klobas}
\affiliation{School of Physics and Astronomy, University of Birmingham, Edgbaston, Birmingham, B15 2TT, UK}
\author{Lorenzo Piroli}
\affiliation{Dipartimento di Fisica e Astronomia, Universit\`a di Bologna and INFN, Sezione di Bologna, via Irnerio 46, I-40126 Bologna, Italy}

\maketitle
	

\prlsection{Introduction} Random ensembles of quantum operators are fundamental tools in quantum information theory~\cite{nielsen2010quantum,wilde2013quantum}, now being routinely employed in several experimental settings such as randomized measurement protocols~\cite{elben2023randomized,elben2019statistsical,cieslinski2024analysing,huang2020predicting} or in the context of quantum advantage demonstrations~\cite{boixo2018characterizing,arute2019quantum,wu2021strong, hangleiter2023computational}.  At the same time, models evolving through discrete random operations provide a powerful computational framework to tackle nonperturbative problems in many-body physics~\cite{fisher2023random,potter2022entanglement}. 

The point of contact is given by the recent identification of quantum circuits --- many-body systems in discrete spacetime --- as useful toy models for generic many-body dynamics~\cite{hayden2007black,bertini2019exact,nahum2017quantum}. Indeed, their minimal structure provides a concrete handle to study hard topical questions pertaining to, \emph{e.g.}, entanglement growth~\cite{nahum2017quantum,zhou2019emergent, bertini2019entanglement, gopalakrishnan2019unitary, zhou2020entanglement, foligno2024quantum, piroli2020exact}, thermalization processes~\cite{piroli2020exact,klobas2021exact,bertini2024east,wang2024exact}, and information scrambling~\cite{hosur2016chaos,vonKeyserlingk2018operator,nahum2018operator, claeys2020maximum, bertini2020scrambling}. Circuit models are also much easier to implement on current quantum devices than continuous-time dynamics~\cite{lloyd1996universal}, therefore provide ideal benchmarking  platforms for current prototypes~\cite{chertkov2022holographic,jones2022small,fischer2024dynamical}. In analogy to random matrix theory, the introduction of randomness simplifies the analytical treatment of circuit models while retaining the basic features of the dynamics.

\begin{figure}[h!]
	\centering
    \includegraphics[width = \linewidth]{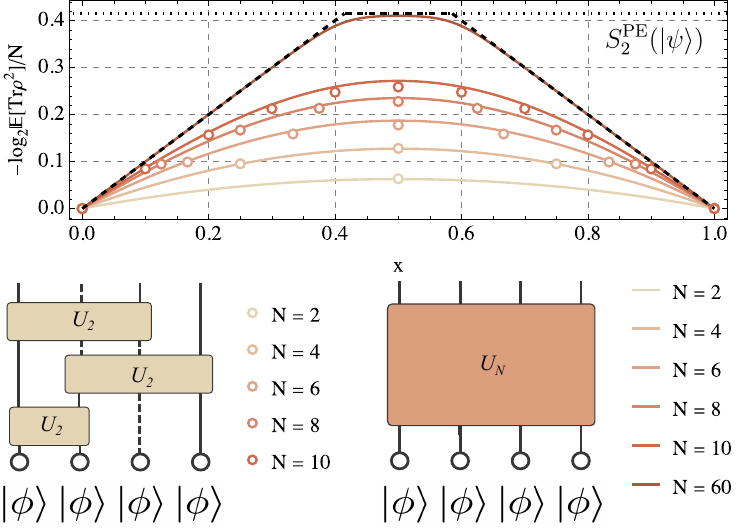}
	\caption{(a) Averaged R\'enyi-$2$ entropies generated by a deep random circuit of two-qubit gates (open markers)  and global random permutations (solid lines) acting on $\ket{\psi}=\ket{\phi(\theta)}^{\otimes N}$, with $\ket{\phi(\theta)}=\cos\theta|0\rangle+\sin(\theta)|1\rangle$ and $\theta = \pi/8$. The black dashed line shows our rigorous bound [Eq.~\eqref{eq:final_bound}] and the black dotted one the second participation entropy $S_{2}^{\rm PE}(\ket{\psi})$. (b) Schematic depiction of the two random circuits considered.}
	\label{fig:results}
\end{figure}

Random circuit models are defined by a few ingredients: the number $k$ of qubits coupled by the local gates, the geometric connectivity of the circuit, and the set of unitaries from which the gates are sampled~\cite{fisher2023random}. In this setting, one is typically interested in the averaged dynamics of quantum correlations, their asymptotic values, and the circuit depths required to reach a stationary output distribution. These features strongly depend on the precise model under consideration: significant theoretical work has recently allowed us to characterize several families of circuits, with systematic results obtained in the case of Haar-random gates~\cite{hunter2019unitary,dalzell2022random,harrow2023approximate,mittal2023local,schuster2024random,laracuente2024approximate}, random Cliffords~\cite{renes2004symmetric,ambainis2007quantum,zhu2016clifford}, or random matchgate circuits~\cite{bauer2019equiliibrium,bernard2021entanglement,dias2021diffusive,bianchi2021page,yu2023free}.

In this work, we consider \emph{random-permutation} circuits, made of gates reshuffling the local computational basis, reproducing the action of reversible classical automata~\cite{aldana_boolean_2003,iadecola2020nonergodic,pizzi2022bridging,iaconis2019anomalous,iaconis2021quantum}. As recently pointed out in Ref.~\cite{bertini2025quantum}, these circuits are appealing because, by acting classically on a given basis of states, they allow one to address the role played by ``quantumness'' in quantum-circuit dynamics. Here, we focus on entanglement, a hallmark of quantum correlations, and present two main results, which appear to be unique features of permutation dynamics. First, we derive upper bounds on the entanglement that can be generated by permutation circuits acting on arbitrary initial states and show that they are generically saturated. This result states that the quantum correlations generated by classical circuits are bounded in terms of some quantum property of the initial state (namely, the degree to which it can be written as a superposition of classical states). Second, starting from simple $N$-qubit states, we compare the averaged R\'enyi-$2$ entropies generated by $(i)$ an infinitely deep random circuit of two-qubit gates and $(ii)$ global random permutations. We show that the two quantities differ for finite $N$, but their Page curves~\cite{page1993average,page1993information} coincide in the thermodynamic limit, see Fig.~\ref{fig:results} for a visual summary. This result highlights an unexpected locality constraint on the set of permutation operators, similar, in spirit, to the one recently established on the set of symmetric unitaries~\cite{marvian2024theory,marvian2024rotationally,marvian2022restrictions}.

\prlsection{Random permutation circuits} We consider a system of $N$ qubits, corresponding to the Hilbert space $\mathcal{H}=\otimes_{j=1}^N \mathcal{H}_j$, with $\mathcal{H}_j\simeq \mathbb{C}^{2}$. We denote by $\sigma_j^\alpha$ the Pauli matrices, with $\alpha=0,1,2,3$ ($\sigma^0=\openone$ being the identity) and by $\{|0\rangle, |1\rangle\}$ the local computational basis.  The permutation operators of interest here are those implementing a reshuffling of the states of the computational basis. Namely, given a computational basis state of $k$ qubits a permutation operator acts as
\be
U^{(\pi)} \ket{s} = \ket{\pi(s)}.
\ee
Here $\pi\in S_{2^k}$ is a permutation of $2^k$ elements, $\ket{s}=\ket{s_1}\otimes \ket{s_2}\cdots\otimes\ket{s_k}$, and $\{s_1,s_2,\ldots,s_k\}$ are the coefficients of the binary representation of $s\in \mathbb Z_{2^k-1}$.

We consider two random-operator ensembles. The first, $\mathcal{E}_{\rm GP}$, is the set of random global permutations and is defined by sampling from a uniform distribution the set of $(2^N)!$ permutation operators acting on $N$ qubits. The second, $\mathcal{E}_{\rm PC}(D)$, corresponds to the set of random permutation circuits of depth $D$. While most of our results will be independent of the circuit connectivity, we focus on non-local circuits made of $2$-qubit gates, which are defined by choosing, at each discrete step, a random pair of qubits $i,j$ and applying a random permutation operator $U^{(\pi)}_{i,j}$ to them. The number of discrete steps is the circuit depth $D$. Following Ref.~\cite{piroli2020random}, we in fact consider a continuous-time version of this model: We associate an interval $\Delta t$ with each discrete step, and apply each random gate $U^{(\pi)}_{i,j}$ with probability $p=N\lambda \Delta t$, where $\lambda>0$. In this way, expectation values of observables computed at time $t$ display a well-defined limit for $\Delta t \to 0$, which allows us to ease some of the analytic computations.  

Given an initial state $\ket{\psi_0}$, we denote by $\ket{\psi_t}$ the state evolved by the circuit up to time $t$, and consider the averaged bipartite R\'enyi-entropies
\begin{equation}
\mathbb{E}[S_\alpha(\rho_A(t))]=\mathbb{E}\left[(1-\alpha)^{-1}\log_2{\rm Tr}(\rho^\alpha_A(t))\right],
\end{equation}
where $\rho_A(t)={\rm Tr}_{\bar{A}}[|\psi_t\rangle\langle\psi_t|]$, $\bar{A}$ is the complement of the region $A$, while $\mathbb{E}[\ldots]$ denotes the ensemble average. The von Neumann entropy is obtained in the limit $\alpha \to 1$.

\prlsection{Entanglement bounds for permutation circuits} Consider a bipartition $\mathcal{H}=\mathcal{H}_A\otimes \mathcal{H}_B$, so that any state can be written as $\ket{\psi_0}=\sum_{n,m}c_{n,m}|n\rangle_A|m\rangle_B$, where $\ket{n}_A$, $\ket{m}_B$ denote basis states for $\mathcal{H}_A$ and $\mathcal{H}_B$. We begin by deriving upper bounds on the R\'enyi-$\alpha$ entropies generated by applying arbitrary permutation circuits to $\ket{\psi_0}$. 

First, we introduce the inverse participation ratio
\begin{equation}\label{eq:inverse_participation}
I_{\alpha}(\ket{\psi})=
  \sum_{s}|\langle s |\psi\rangle|^{2\alpha}\,,
\end{equation}
where $\{\ket{s}\}$ is the computational basis of $\mathcal{H}$. $I_{\alpha}(\ket{\psi})$ is closely related to the so-called participation entropy (PE) $S^{\rm PE}_\alpha(\ket{\psi})=(1-\alpha)^{-1}\log_2 I_{\alpha}(\ket{\psi})$~\cite{stephan2009shannon,stephan2009renyi,alcaraz2013,stephan2014renyi,luitz2014participation}. We also introduce the overlap 
\begin{equation}\label{eq:overlap_z}
    z(\ket{\psi})=|\langle  a_N|\psi\rangle|\,,
\end{equation}
where we denoted by $\ket{a_l}=|+\rangle^{\otimes l}=(\ket{0}+\ket{1})^{\otimes l}/2^{l/2}$ the ``maximally anti-localized'' state of $l$ qubits. The state $\ket{a_N}$ can be easily seen to maximize the PE. Crucially, $z(\ket{\psi})$ and $I_{\alpha}(\ket{\psi})$ are invariant under any operator that permutes the computational basis (therefore are time independent) and the idea is to bound $S_\alpha(\rho_A(t))$ in terms of these quantities. 
Such a bound is also consistent with the numerical observations of Ref.~\cite{pizzi2022bridging}.

To proceed we consider the initial reduced density matrix $\rho_A(0)=\sum_{n,m,p}c_{n,m}c^\ast_{p,m}|n\rangle_A\langle p|$ and the dephasing channel $\mathcal{D}_l(\cdot)=\sum_{s}\bra{s}\cdot \ket{s}\, \ket{s}\!\bra{s}$, where $\{\ket{s}\}$ denotes the computational basis of $\otimes_{j=1}^{l}\mathcal{H}_j$. Since the dephasing channel is unital, $\mathcal{D}_{l}(\openone)=\openone$, the data-proccessing inequality for the sandwiched R\'enyi divergences~\cite{frank2013monotonicity} implies ${S_\alpha(\rho_A)\leq S_{\alpha}(\mathcal{D}_{|A|}(\rho_A))}$. Next, we note $S_{\alpha}(\mathcal{D}_{|A|}(\rho_A))=(1-\alpha)^{-1}\log_2 \sum_{n}p_n^\alpha$, where ${p_n=\sum_m |c_{n,m}|^2}$. Finally, using that the (classical) R\'enyi entropies of marginal probability distributions are bounded by the R\'enyi entropy of the joint distribution~\cite{linden2013structure}, we obtain
\begin{equation}\label{eq:s_a_vs_pe}
S_{\alpha}(\rho_A(t))\leq S^{\rm PE}_\alpha(|\psi_0\rangle)\,.
\end{equation}

Eq.~\eqref{eq:s_a_vs_pe} provides a bound that is typically not tight. For instance, when initializing the system in the state $\ket{\psi_0}=\ket{a_N}$, the R\'enyi entropies remain identically zero, while $S^{\rm PE}_\alpha(|\psi_0\rangle)$ scales linearly in $N$. This remark suggests improving the bound by taking into account the overlap $z(|\psi_0\rangle)$. To this end, we introduce the dephasing channel in the $x$-basis, $\widetilde{\mathcal{D}}_l(\cdot)=\sum_{s}\bra{\nu_s} \cdot \ket{\nu_s} \ket{\nu_s}\!\bra{\nu_s}$, where $\ket{\nu_s}=(\sigma^z_1)^{s_1}\cdots (\sigma^z_l)^{s_l}|a_l\rangle$, and $s_1,\ldots ,s_l$ are the digits in the binary representation of $s$. Using again the data-processing inequality ($\widetilde{\mathcal{D}}_l$ is also unital), we obtain $S_\alpha(\rho_A)\leq S_{\alpha}(\widetilde{\mathcal{D}}_{|A|}(\rho_A))$, and combining with $\sum_s \braket{\nu_s|\rho_A|\nu_s}^\alpha \geq \braket{a_{|A|}|\rho_A|a_{|A|}}^{\alpha}\geq z^{2\alpha}$, we arrive at $S_{\alpha}(\rho_A(t))\leq (1-\alpha)^{-1}\log_2 z^{2\alpha}$. Therefore, putting everything together, we obtain 
\begin{equation}\label{eq:final_bound}
S_{\alpha}(\rho_A(t))\leq {\rm min}[|A|, S^{\rm PE}_\alpha(|\psi_0\rangle), (1-\alpha)^{-1} \log_2 z^{2\alpha}]\,,
\end{equation}
where we also took into account that the R\'enyi entropy cannot exceed $|A|$, the number of qubits in $A$ (assuming $|A|\leq N/2)$. Eq.~\eqref{eq:final_bound} is the first main result of our work.

A few remarks are in order. First, we stress that the intermediate result, Eq.~\eqref{eq:s_a_vs_pe}, already appeared in Ref.~\cite{gopalakrishnan2018facilitated}, although its derivation there was not fully formalized, cf. also Ref.~\cite{feng2025dynamics}. Second, we note that the quantities appearing in the right-hand side of Eq.~\eqref{eq:final_bound} can be efficiently measured in families of low-entangled, short-correlated states. For instance, both the overlap and the PE may be obtained applying the method of Ref.~\cite{vermersch2025many} (see also the recent proposal~\cite{chen2025zipping} for measuring the von Neumann PE). Third, we note that $(1-\alpha)^{-1} \log_2 z^{2\alpha}\to \infty$ for $\alpha\to 1$, thus trivializing the $z$-dependence of the upper bound. Still, using the Fannes inequality~\cite{audenaert2007sharp}, we are able to derive the following alternative bound for $\alpha=1$~\cite{SM}
\begin{equation}\label{eq:bound_von_Neumann}
S_{1}(t)\leq {\rm min}\left[|A|, S^{\rm PE}_1(|\psi\rangle),{|A|}\sqrt{1-z^2}+O(1)\right]\,.
\end{equation}
Finally, an important question is whether the bound is tight, as in principle there could be other missing dynamical invariants. In the rest of this Letter we provide analytical and numerical evidence that, for \emph{typical} permutation operators, the bound in Eq.~\eqref{eq:final_bound} for the R\'enyi-$2$ entropy is indeed saturated at the leading order in $N$. 

To this end, we now proceed to compute the averaged bipartite entanglement in the global permutation [$\mathcal{E}_{\rm GP}$] and the permutation-circuit [$\mathcal{E}_{\rm PC}(t)$] ensembles, respectively. We are interested in two aspects: estimating the time needed for averages over $\mathcal{E}_{\rm PC}(t)$ to approach stationary values, and understanding if the entanglement averaged over $\mathcal{E}_{\rm PC}(\infty)$ and $\mathcal{E}_{\rm GP}$ coincide. These problems are well understood in the Haar random case~\cite{harrow2023approximate,dalzell2022random,mittal2023local,schuster2024random,laracuente2024approximate}: sufficiently connected Haar random circuits over $N$ qubits approach the ensemble of global Haar random operators at infinite depths, while for $2$-local circuits, the entanglement approaches its asymptotic value at the \emph{scrambling time} $t_s\sim O(\log N)$~\cite{hayden2007black,sekino2008fast,lashkari2013towards}. In the case at hand, however, the situation is much less clear as two-qubit permutation gates belong to the Clifford group~\cite{gottesman1997stabilizer,gottesman1998theory} and circuits generated by Clifford gates are well known to display special features (including being efficiently simulable when initialized in so-called stabilizer states~\cite{nielsen2010quantum}).

\prlsection{Effective dynamics in replica space} As pointed out in Refs.~\cite{iaconis2019anomalous,iaconis2021quantum}, several features of permutation circuits can be simulated by sampling the basis states appearing in the decomposition of the input state and computing the corresponding classical evolution~\cite{iaconis2019anomalous}. The computational cost to simulate the expectation value of an observable is then determined by the statistical variance of the sampled values. While this method is provably efficient for observables with bounded operator norm, entanglement-related quantities require a number of samples growing exponentially in $N$, thus offering no real advantage over a brute-force state-vector simulation of the quantum dynamics~\cite{SM}. Therefore, studying entanglement dynamics under permutation circuits is nontrivial.

Here we present an analytic approach to compute the purity $\mathbb{E}[2^{S_2(t)}]$~\footnote{We note that the logarithm of the averaged purity is not the averaged Rényi-$2$ entropy, as the disorder average is taken inside the logarithm. However, when $N\to\infty$ the fluctuations become small, so that the logarithm of the averaged purity is a good approximation for the averaged R\'enyi-$2$ entropy.}, which makes use of several recently developed tools for random quantum circuits~\cite{fisher2023random}. We start by exploiting the Choi-Jamiolkowski isomorphism~\cite{nielsen2010quantum} to map the operator $\rho_A(t)\otimes \rho_A(t)$ to the state
\begin{equation}
\label{eq:Choi_representation}
    |\rho_A(t)\otimes \rho_A(t)\rangle=\openone \otimes \rho_A(t)\otimes \openone\otimes \rho_A(t)|\mathcal{I}\rangle\,,
\end{equation}
where $|\mathcal{I}\rangle=|I^+\rangle^{\otimes N}\in \mathcal{H}^{\otimes 4}$ and $\ket{I^+}=\sum_{a,b=0}^1|aabb\rangle$. The purity of $\rho_A(t)$ can then be written as~\cite{SM}
\begin{equation}
    \text{Tr}_A\mathbb{E}[\rho^2_A(t)]=\bra{G_{|A|,N-|A|}}\rho_A(t)\otimes \rho_A(t)\rangle\,,
\end{equation}
where we have introduced the boundary state $|G_{n_{-},n_{+}}\rangle=\ket{I^-}^{\otimes n_-}\otimes\ket{I^+}^{\otimes n_+}$, with $\ket{I^+}$ defined above and $\ket{I^-}=\sum_{a,b=0}^1|abba\rangle$. This formalism is useful because it allows one to reformulate the computation of the time-dependent purity in terms of the dynamics of the averaged vector $\mathbb{E}\left[\ket{\rho(t)\otimes\rho(t)}\right]$. Exploiting now the fact that the randomness is uncorrelated in space and time, a standard derivation shows~\cite{SM}
\begin{equation}
\label{eq:diff_eq}
    \frac{d}{dt}\mathbb{E}\left[\ket{\rho(t)\otimes\rho(t)}\right] = - \mathcal{L}\mathbb{E}\left[\ket{\rho(t)\otimes\rho(t)}\right]
\end{equation}
where
\begin{equation}
    \label{eq:Lindbladian_continuum_def_main}
    \mathcal{L} = \frac{2\lambda}{N-1}\sum_{\langle j, k\rangle}(1-\mathcal{U}_{j,k})\,.
\end{equation}
Here the sum is over sites that are nearest neighbors according to the circuit connectivity, we set $\mathcal{U}_{j,k} = \mathbb{E}[U^{*}_{j,k}\otimes U_{j,k} \otimes U^{*}_{j,k} \otimes U_{j,k}]$, and $(\cdot)^\ast$ denotes complex conjugation. By explicit computation~\cite{bertini2025quantum}, the average $\mathcal{U}_{j,k}$ is seen to be a non-unitary operator acting on the tensor product of local subspaces of dimension $d_{\rm eff}=15$. Eq.~\eqref{eq:diff_eq} describes a deterministic evolution in a \emph{replica} Hilbert space of local dimension $d_{\rm eff}$. Although it does not require sampling over disorder realizations, its solution is still challenging due to the exponential growth of the Hilbert-space dimension. In the following, we simplify the problem relying on two key properties.

First, for the non-local circuits studied in our work, the sum in Eq.~\eqref{eq:Lindbladian_continuum_def_main} runs over arbitrary pairs of sites so that $\mathcal{L}$ is invariant under site permutation. This introduces a significant simplification~\cite{sunderhauf2019quantum,piroli2020random}: when starting from permutation-symmetric initial conditions, the dynamics takes place in the symmetric subspace, whose dimension is $D_{\rm symm}=\binom{N+d_{\rm eff}-1}{d_{\rm eff}-1}$. Unfortunately, despite growing only polynomially in $N$, it does so with a large exponent, $D_{\rm symm}~\sim N^{14}$, thus making the direct numerical solution of Eq.~\eqref{eq:Lindbladian_continuum_def_main} unfeasible for large $N$. To make progress, we exploit the fact that two-qubit permutation gates belong to the Clifford group. This property allows us to further simplify the problem and obtain analytic results for certain families of initial states, using a method that is reminiscent of the twirling technique of Ref.~\cite{sunderhauf2018localization}.

\prlsection{Exact dynamics from typical initial states}  We consider random product states $\ket{\psi}=\bigotimes_j \ket{\phi_j}$, where ${\ket{\phi_j}=v_{j}\ket{0}}$ with $v_j$ Haar random unitaries. Denoting by $\mathcal{P}_N$ the set of Pauli strings without overall phases, the invariance of the Haar measure under left multiplication implies that $(P^\ast\otimes P)^{\otimes 2}$ leaves $\mathbb{E}_{\{\phi_j\}}[\otimes_{j}(|\phi_j\rangle^\ast|\phi_j\rangle)^{\otimes 2}]$ invariant for any $P$ in $\mathcal{P}_N$. Therefore, also 
\begin{equation}\label{eq:defPauliProjector}
 \!\!\!\frac{1}{4^N} \!\!\! \sum_{P\in \mathcal{P}_N} \!\!(P^\ast\otimes P)^{\otimes 2} \!=\! 
\left  [\frac{1}{4}\! \sum_{\alpha=0}^3\sigma_j^{\alpha \ast} \!\otimes \sigma_j^\alpha \!\otimes \sigma_j^{\alpha \ast} \!\otimes \sigma_j^\alpha\right ]^{\!\!\otimes N} \!\!\!\!\!,
\end{equation}
leaves invariant the averaged state. Denoting now by $\Pi_j$ the quantity in brackets on the r.h.s., we note that $\Pi_j$ is a projector onto a $4$ dimensional subspace of $\mathcal{H}_j^{\otimes 4}$. Therefore the averaged state is left invariant by the projector $(\otimes_j \Pi_j)$. Moreover, we also have $(\otimes_j \Pi_j) {\mathcal{L}}= \mathcal{L}(\otimes_j \Pi_j)$, which follows from the fact that the two-qubit gates are Clifford operators~\cite{SM}. Putting all together and setting $\tilde{\mathcal{L}}\equiv (\otimes_j \Pi_j) \mathcal{L}(\otimes_j \Pi_j)$, we obtain that the solution to Eq.~\eqref{eq:diff_eq} after averaging over the random initial states can be written as
\begin{equation}\label{eq:projected_dynamics_main}
\ket{\rho(t)\otimes\rho(t)}=e^{-\mathcal{\tilde{L}}t}\mathbb{E}_{\phi_j}[\otimes_{j}\Pi_j|(|\phi_j\rangle^\ast|\phi_j\rangle)^{\otimes 2}\rangle]\,.
\end{equation}
That is, the dynamics takes place in the site-permutation symmetric subspace of $\tilde{\mathcal{H}}=\otimes_j\tilde{\mathcal{H}}_j$ with ${\rm dim}(\tilde{\mathcal{H}}_j)=4$. The corresponding dimension is now $\binom{N+3}{3}\sim N^3$, which allows one to perform efficient numerical computations.

In fact, Eq.~\eqref{eq:projected_dynamics_main} allows us to go further and derive a system of differential equations encoding the purity dynamics. Specifically, we set $G_{p,q,r,s}= \bra{G_{p,q,r,s}}\rho_A(t)\otimes \rho_A(t)\rangle$, where $|G_{p,q,r,s}\rangle=\ket{I^-}^{\otimes p}\!\otimes\! \ket{I^+}^{\otimes q} \!\otimes\! \ket{I^0}^{\otimes r} \!\otimes\! \ket{I^x}^{\otimes s}$ with $\ket{I^\pm}$ are defined above, while
$\ket{I^0}=\sum_{a=0}^1\ket{aaaa}$, $\ket{I^x}=\sum_{a,b=0}^1\ket{abab}$. Using Eq.~\eqref{eq:projected_dynamics_main} we  find~\cite{SM}
\begin{equation}\label{eq:differential_system_main}
  \frac{dG_{p,q,r,s}(t)}{dt}=\smashoperator{\sum_{\{\ell_k\}_{k=1}^4}}
  c[\{\ell_k\}]G_{p+\ell_1,q+\ell_2,r+\ell_3,s+\ell_4}(t)\,.
\end{equation}
The coefficients $c[\{\ell_k\}]$ are reported explicitly in Ref.~\cite{SM}, together with an expression for the initial condition $G_{p,q,r,s}(0)$ corresponding to random initial product states. Eq.~\eqref{eq:differential_system_main} is the second main result of our work. 

\prlsection{Entanglement dynamics and Page curves} 
Before illustrating the solution to Eq.~\eqref{eq:differential_system_main}, we complete our analysis by discussing the averaged bipartite purity over the global ensemble $\mathcal{E}_{\rm GP}$. This computation can be performed via a similar replica-space approach, but it is now much simpler due to the absence of spatial structures and time dependence. In this case, we can perform the computation for arbitrary initial product states, obtaining an explicit, although unwieldy, formula~\cite{SM}. For instance, the averaged purity for random initial product states reads
\begin{equation}\label{eq:purity_global_average}
\!\!\mathbb{E}_{\rm GP}[{\rm Tr}\rho^2_A]\!=\!
  \frac{\!y (\xi^{1-x}\!+\!\xi^x)q \!-\![(\xi\! -\! \xi^{x})(\xi^{1-x}\!-\!\xi)+2y]p}{y (\xi-3) (\xi-2) (\xi-1)}
\end{equation}
where $p=7 + (-4 + \xi) \xi$, $q=8 + (-5 + \xi) \xi$, $\xi=2^N$ and $y=3^N$. As another example, Fig.~\ref{fig:results} reports the evaluation of the analytic formula for an initial state 
\begin{equation}\label{eq:prod_state_theta}
    \ket{\psi_0}=\ket{\phi(\theta)}^{\otimes N},\quad \ket{\phi(\theta)}=\cos\theta|0\rangle+\sin(\theta)|1\rangle\,.
\end{equation}

 \begin{figure}
	\centering
    \includegraphics[width = \linewidth]{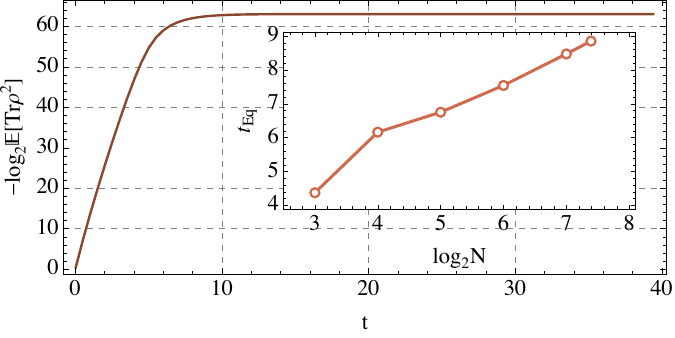}
	\caption{R\'enyi-$2$ entropy dynamics for random-permutation circuits and random initial product states. The system size is $N=128$. Inset: the equilibration time, as a function of $N$.}
	\label{fig:time_evolution}
\end{figure}

As we vary the subsystem size $|A|=xN$ in the limit ${N\to\infty}$, $x\in (0,1)$, the logarithm of the averaged purity yields the R\'enyi-$2$ entropy Page curve corresponding to the ensemble $\mathcal{E}_{\rm GP}$~\footnote{We have again assumed $\log(\mathbb{E}_{\rm GP}[{\rm Tr}\rho^2_A])\sim\mathbb{E}_{\rm GP}(\log {\rm Tr}\rho^2_A])$}. We have derived the infinite-$N$ Page curve for an arbitrary product-state of the form \eqref{eq:prod_state_theta}: Setting $\bar{s}_2(x)=\lim_{N\to\infty}( -\log(\mathbb{E}_{\rm GP}[{\rm Tr}\rho^2_A]))/N$, and choosing $x\leq 1/2$, we find~\cite{SM}
\begin{equation}\label{eq:s2bar}
\bar{s}_2(x)={\rm min}\{x,s_2^{\rm PE}(\theta),-2\log_2 \zeta^2\}\,,
\end{equation}
where $s_2^{\rm PE}=\lim_{N\to\infty}S_2^{\rm PE}(|\psi\rangle)/N$ and $\zeta^2=|\braket{\phi(\theta) |+}|^2$. Note that, since $-2\log_2 \zeta^2=\lim_{N\to\infty}[-\log_2(z^{2})/N]$ with $z=|\braket{\psi|a_N}|$, Eq.~\eqref{eq:s2bar} shows that the bipartite entanglement saturates the bound in Eq.~\eqref{eq:final_bound} as $N\to \infty$. 

\begin{figure}
	\centering
    \includegraphics[width = \linewidth]{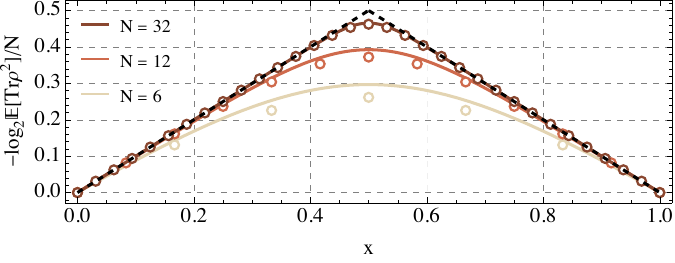}
	\caption{Comparison between the Page curves obtained averaging over the late-time circuit ensemble, $\mathcal{E}_{\rm PC}(\infty)$ (open markers), and the global-permutation ensemble, $\mathcal{E}_{\rm GP}$ (solid lines). The system is initialized in a random product state. The upper bound is shown in black dashed lines.}
	\label{fig:page_curves}
\end{figure}

We finally discuss the entanglement dynamics of the permutation circuit, as described by Eq.~\eqref{eq:differential_system_main}. Fig.~\ref{fig:time_evolution} shows the time evolution from random initial states. Based on a scaling analysis perfomed at increasing system sizes, we see that the time needed for the entanglement to equilibrate to its maximum value scales as $t_e\sim O(\log N)$, similarly to the scrambling time in Haar random circuits. Next, we compare the averaged purity computed by solving Eq.~\eqref{eq:differential_system_main} up to large times and the one corresponding to global random permutations [cf.~Eq.~\eqref{eq:purity_global_average}]. Our results are shown in Fig.~\ref{fig:page_curves}. The steady-state solution of Eq. \eqref{eq:differential_system_main} is obtained by solving the system $d G_{p,q,r,s}(t)/dt=0$, and the exact expression of the averaged purity is reported in \cite{SM}. Comparing the latter with \eqref{eq:purity_global_average} shows that the finite-$N$ results are different, whereas the Page curves coincide in the thermodynamic limit. One can repeat the stationary-state analysis also for arbitrary initial states, which are not invariant under~\eqref{eq:defPauliProjector}. Interestingly, in this case the difference between the two Page curves does not necessarily vanish in the $N\to\infty$ limit~\cite{SM}.

We stress that, although our gates belong to the Clifford group, here we are applying the circuit to arbitrary (random) initial states, yielding non-stabilizer states for which entanglement dynamics is typically hard to simulate~\cite{jozsa2013classical, yoganathan2019quantum}. Therefore, it is not a priori obvious that the entanglement Page curves corresponding to $\mathcal{E}_{\rm GP}$ and $\mathcal{E}_{\rm PC}(\infty)$ should differ for finite $N$, and only coincide in the thermodynamic limit. A natural question, however, is what happens if we break the Clifford property. This can be easily achieved by increasing the support $k$ of our local permutation gates or introducing phases. 

\prlsection{$k$-local permutation and random-phase gates} Permutation gates are typically not Clifford for $k\geq 3$. In fact, it is known that circuits made of random $3$-qubit gates form permutation $t$-designs for all values of $t$~\cite{gowers1996almost,chen2024incompressibility,gay2025pseudorandomness}. That is, for sufficiently connected and deep circuits, all the moments of the permutation-circuit ensemble match those of the ensemble of global permutations taken from the \emph{alternating group}, which is formed by the even permutations. At the same time, the moments of the alternating and full permutation groups coincide up to order $t= 2^N-2$~\cite{chen2024incompressibility}. Then, since the full entanglement spectrum of the reduced density matrix is uniquely determined by its R\'enyi-$\alpha$ entropies with integer $\alpha\leq 2^{N/2}$, the entanglement Page curves corresponding to the infinite-time circuit ensemble and $\mathcal{E}_{\rm GP}$ coincide for all $N$.

Going beyond permutations, one can consider more general unitaries that preserve the computational basis, by allowing for additional phases setting $U^{(\pi)}_{ij}\ket{s_1, s_2}=e^{i\varphi_{s_1,s_2}}\ket{\pi(s)_1, \pi(s)_2}$. Circuits made of such gates were studied in Ref.~\cite{iaconis2019anomalous,iaconis2021quantum}, where they were referred to as automaton circuits. We can define automaton-circuit ensembles that are analogous to $\mathcal{E}_{\rm PC}(t)$ and $\mathcal{E}_{\rm GP}$. The first one is obtained by considering a circuit of random $2$-qubit gates, where the permutations and phases are randomly drawn from the uniform measure. As before, one can then take a continuous-time limit, defining the ensemble $\mathcal{E}_{\rm PPC}(t)$. The second ensemble, $\mathcal{E}_{\rm GPP}$, is constructed by applying global random unitaries acting as $U^{(\pi)}\ket{s}=e^{i\varphi_{s_1\ldots s_N}}\ket{\pi(s)}$, with random $\pi$ and $\varphi$.

In this setting, it can be checked that the maximally antilocalized state $\ket{a_N}$ is not conserved by the circuit, so that the bipartite entanglement can only be bounded by the PE as in Eq.~\eqref{eq:s_a_vs_pe}. Similarly to the case of permutations, one can ask if the bound is tight and the entanglement entropies averaged over $\mathcal{E}_{\rm GPP}$ and $\mathcal{E}_{\rm PPC}(t)$ coincide. These questions can be addressed following the steps described above for the random permutations. It turns out that the random phases introduce great simplifications, allowing us to arrive at explicit results for very general initial states, including higly entangled states~\cite{SM}. Our analysis shows that the entanglement entropies averaged over $\mathcal{E}_{\rm PPC}(\infty)$ and $\mathcal{E}_{\rm GPP}$ coincide for arbitrary $N$, and that, similarly to the case of random permutations, the bound in Eq.~\eqref{eq:s_a_vs_pe} is saturated in the thermodynamic limit.

\prlsection{Outlook} Our work raises several questions. First, it is important to understand if the bound obtained for the von Neumann entropy, Eq.\eqref{eq:bound_von_Neumann}, is also typically tight. Because of the different functional dependence on $z$ compared to Eq.~\eqref{eq:final_bound}, answering this question in the affermative would reveal an interesting phenomenology: For initial states close to $\ket{a_N}$, the R\'enyi Page curves corresponding to $\mathcal{E}_{\rm GP}$ would be all vanishing, while the von Neumann Page curve would not. Moreover, further investigation should be devoted to understanding which types of initial states, beyond random ones, saturate the bound \eqref{eq:final_bound} in the thermodynamic limit. Next, a natural direction would be to consider adding minimal ingredients to the permutation circuit, exploring how the late-time output distribution is modified. For instance, by adding measurements, a natural question is to characterize the so-called projected ensemble~\cite{cotler2023emergentquantum,choi2023preparing} arising from permutation-circuit dynamics.

\prlsection{Acknowledgments} This work was funded by the European Union (ERC, QUANTHEM, 101114881) (D.\ S.-S., M.\ M., and L.\ P.). Views and opinions expressed are however those of the author(s) only and do not necessarily reflect those of the European Union or the European Research Council Executive Agency. Neither the European Union nor the granting authority can be held responsible for them. We also acknowledge financial support from the Royal Society through the University Research Fellowship No.\ 201101 (B.\ B.) and the Leverhulme Trust through the Early Career Fellowship No.\ ECF-2022-324 (K.\ K.). 


\let\oldaddcontentsline\addcontentsline
\renewcommand{\addcontentsline}[3]{}
\bibliography{bibliography}
\let\addcontentsline\oldaddcontentsline
\onecolumngrid
\newpage

\appendix
\setcounter{equation}{0}
\setcounter{figure}{0}
\renewcommand{\thetable}{S\arabic{table}}

\renewcommand{\theequation}{S\thesection.\arabic{equation}}

\setcounter{defn}{0}
\setcounter{thm}{0}
\setcounter{figure}{0}

\setcounter{secnumdepth}{2}

\begin{center}
	{\Large \bf Supplemental Material}
\end{center}

Here we provide additional details on the results stated in the main text.

\tableofcontents

\section{Preliminaries}
 In this SM, we employ the Choi-Jamiolkowski isomorphism \cite{nielsen2010quantum} to write a state $\rho(t) =\ket{\psi(t)}\bra{\psi(t)}$ in a 2-replica space as
\begin{equation}    
\label{eq:Choi_representation_SM}
|\rho(t)\otimes \rho(t)\rangle=\openone \otimes \rho(t)\otimes \openone\otimes \rho(t)|\mathcal{I}\rangle\,,
\end{equation}
where $|\mathcal{I}\rangle=|I^+\rangle^{\otimes N}\in \mathcal{H}^{\otimes 4} \simeq (\mathbb{C}^q)^{\otimes N}$ and $\ket{I^+}=\sum_{a,b=0}^{q-1}|aabb\rangle$ is the maximally entangled state in the 2-replica space. We will also make use of the following states:
\begin{equation}
\label{eq:I_states_definition_SM}
\ket{I^-}= \sum_{a,b=0}^{q-1} \ket{abba}, \quad \ket{I^x}= \sum_{a,b=0}^{q-1} \ket{abab}, \quad \ket{I^0}= \sum_{a=0}^{q-1} \ket{aaaa}.
\end{equation}

To obtain the continuous-time evolution of the state \eqref{eq:Choi_representation_SM}, we follow a simplified version of the model in \cite{piroli2020random}. Namely, at every discrete time step $t$, we apply a two-body gate $U_{j,k}$, $1\le j  < k \le N$, with a certain probability $p$, so that at time $t+\Delta t$ the state is:
\begin{equation}
\ket{\rho(t+\Delta t)\otimes \rho(t+\Delta t)} = \openone \otimes U(t+\Delta t)\rho(t)U^\dagger(t + \Delta t) \otimes \openone \otimes U(t+\Delta t)\rho(t)U^\dagger(t + \Delta t) \ket{I^+},
\end{equation}
where $U(t+\Delta t) = U_{j,k} U(t)$ with probability $p= \lambda N \Delta t$, $\lambda > 0$, and $U(t+\Delta t) = U(t)$ with probability $1-p$. Then we average over the sites $j,k$ and over the ensemble $\mathcal{E}$ from which the operators $U_{j,k}$ are drawn. After taking the limit $\Delta t \to 0$, this yields a Lindblad equation for the state $\mathbb{E}[\ket{\rho(t)\otimes\rho(t)}]$:
\begin{equation}
\label{eq:Lindblad_equation_general_SM}
  \frac{d}{dt}\mathbb{E}\left[\ket{\rho(t)\otimes\rho(t)}\right] = - \mathcal{L}\mathbb{E}\left[\ket{\rho(t)\otimes\rho(t)}\right],
\end{equation}
with
\begin{equation}
    \label{eq:Lindbladian_continuum_def_SM}
    \mathcal{L} = \frac{2\lambda}{N-1}\sum_{1\leq j< k\leq N}(1-\mathcal{U}_{j,k}),\quad \mathcal{U}_{j,k} = \mathbb{E}\left[U^{*}_{j,k}\otimes U_{j,k} \otimes U^{*}_{j,k} \otimes U_{j,k}\right].
\end{equation}

The ensemble-averaged purity $\text{Tr}_A\mathbb{E}[\rho_A^2(t)]$ of the state $\ket{\rho(t) \otimes \rho(t)}$ in a bipartition of the total system, $\mathcal{S}=A \cup \bar{A}$. is obtained by introducing a swap operator $X_A$ which exchanges the two copies of the subsystem $A$. In the Choi-Jamiolkowki representation, the swap operator reads:
\begin{equation}
\ket{X_A} = \bigotimes_{i \in A}\ket{I^-}_i \bigotimes_{j \in \bar{A}} \ket{I^+}_j,
\end{equation}
and the averaged purity is given by the overlap 
\begin{equation}
\text{Tr}_A\mathbb{E}[\rho_A^2(t)] = \text{Tr}\{X_A \mathbb{E}[\rho(t)\otimes \rho(t)]\} = \braket{X_A|\mathbb{E}[\rho(t)\otimes \rho(t)]}.
\end{equation}
We will consider two ensembles: random permutation circuits, $\mathcal{E}_\text{PC}$, and circuits made of random permutation and random phase gates, $\mathcal{E}_\text{PPC}$. When the matrices $U_{i,j}$ are drawn from these two ensembles, the averaged two-body gates $\mathcal{U}_{i,j}$ are operators acting on (tensor products of) replica Hilbert spaces of small local dimension $d_\text{eff}$. Namely, in the case of $\mathcal{E}_\text{PPC}$ the space $\mathcal{H}_\text{eff}$ is spanned by tensor products of the states $\ket{I^+}_i,\, \ket{I^-}_i,\,\ket{I^0}_i$, while for the ensemble $\mathcal{E}_\text{PC}$, when the local dimension is $q=2$ and the initial state is a random product state, $\mathcal{H}_\text{eff}$ is spanned by tensor products of $\ket{I^+}_i,\, \ket{I^-}_i,\,\ket{I^x}_i,\,\ket{I^0}_i$.

\subsection{Local dimensional reduction for $2$-qubit permutation gates}
\label{sec:local dimensional reduction}

 In this section, we show that, for random-permutation circuits, the averaged $2$-replica dynamics from random-product states takes place in an effective Hilbert space of dimension $4$.

We consider random product states
\begin{equation}
\ket{\psi}=\bigotimes_j \ket{\phi_j}\,,    
\end{equation}
where $\ket{\phi_j}=v_{j}\ket{0}$ and $v_j$ are independent Haar random single-qubit unitaries. Next, we denote by $\tilde{\mathcal{P}}_N$ the set of all $N$-qubit Pauli strings with an overall phase $\phi\in [\pm 1,\pm i]$, and by $\mathcal{P}_N$ the set of all Pauli strings without overall phases. The invariance of the Haar measure under left multiplication implies 
\begin{equation}
(P^\ast\otimes P)^{\otimes 2}\mathbb{E}_{\{\phi_j\}}[\otimes_{j}(|\phi_j\rangle^\ast|\phi_j\rangle)^{\otimes 2}]=\mathbb{E}_{\{\phi_j\}}[\otimes_{j}(|\phi_j\rangle^\ast|\phi_j\rangle)^{\otimes 2}]\,,   
\end{equation}
where $\mathbb{E}_{\{\phi_j\}}[\cdot]$ denotes averaging over the initial random product states. Since all $2$-qubit permutation gates $U$ are in the Clifford group, we have
\begin{equation}
    U^\dagger PU=\alpha Q\,,
\end{equation}
where $Q\in \mathcal{P}_N$, $|\alpha|=1$, and so
\begin{equation}
    (U^\dagger)^\ast P^\ast U^\ast=\alpha^\ast Q^\ast\,.
\end{equation}
These equations imply
\begin{equation}
(P^\ast\otimes P)^{\otimes 2}(U^\ast\otimes U\otimes U^\ast\otimes U)=  (U^\ast\otimes U\otimes U^\ast\otimes U^\ast) (Q^\ast\otimes Q)^{\otimes 2}\,,
\end{equation}
and so finally
\begin{equation}\label{eq:identity_average}
    \frac{1}{4^N}\sum_{P\in \mathcal{P}_N}(P^\ast\otimes P)^{\otimes 2}( U^\ast\otimes U\otimes U^\ast\otimes U)=  (U^\ast\otimes U\otimes U^\ast\otimes U) \frac{1}{4^N}\sum_{P\in \mathcal{P}_N}(P^\ast\otimes P)^{\otimes 2}\,.
\end{equation}

On the other hand, 
\begin{equation}
\label{eq:projector_Clifford_SM}
\frac{1}{4}\sum_{\alpha=0}^3(\sigma_j^\alpha)^\ast\otimes \sigma_j^\alpha \otimes (\sigma_j^\alpha)^\ast \otimes \sigma_j^\alpha=\Pi_j\,,
\end{equation}
where $\Pi_j$ is a projector onto a $4$ dimensional subspace of $\mathcal{H}_j^{\otimes 4}$, so that $\mathbb{E}_{\{\phi_j\}}[\otimes_{j}(|\phi_j\rangle^\ast|\phi_j\rangle)^{\otimes 2}]=\mathbb{E}_{\{\phi_j\}}[\otimes_{j}\Pi_j(|\phi_j\rangle^\ast|\phi_j\rangle)^{\otimes 2}]$. Using now Eq.~\eqref{eq:identity_average} we have
\begin{align}
 \mathbb{E}_{\{\phi_j\}}[(1-\mathcal{L})^{k}\otimes_{j}(|\phi_j\rangle^\ast|\phi_j\rangle)^{\otimes 2}]&=   \left[(1-\mathcal{L})^{k}\frac{1}{4^N}\sum_{P\in \mathcal{P}_N}(P^\ast\otimes P)^{\otimes 2}
 \otimes_{j}(|\phi_j\rangle^\ast|\phi_j\rangle)^{\otimes 2} \right]\nonumber\\
 &= \left[(1-\mathcal{L})^{k-1}\left(\frac{1}{4^N}\sum_{P\in \mathcal{P}_N} (P^\ast\otimes P)^{\otimes 2}\right)(1-\mathcal{L})\left(\frac{1}{4^N}\sum_{P\in \mathcal{P}_N} (P^\ast\otimes P)^{\otimes 2}\right)
 \otimes_{j}(|\phi_j\rangle^\ast|\phi_j\rangle)^{\otimes 2}\right]\nonumber\\
 &=\mathbb{E}_{\{\phi_j\}}[(1-\mathcal{\tilde{L}})^{k}\otimes_{j}\Pi_j(|\phi_j\rangle^\ast|\phi_j\rangle)^{\otimes 2}],
\end{align}
where $\tilde{\mathcal{L}}=(\otimes_j \Pi_j)\mathcal{L}(\otimes_j \Pi_j)$. Putting all together, we obtain the following solution to Eq.~\eqref{eq:diff_eq}, after averaging over the random initial states:
\begin{equation}\label{eq:projected_dynamics}
\ket{\rho(t)\otimes\rho(t)}=e^{-\mathcal{\tilde{L}}t}\mathbb{E}_{\phi_j}[\otimes_{j}\Pi_j(|\phi_j\rangle^\ast|\phi_j\rangle)^{\otimes 2}\,.
\end{equation}
That is, the dynamics takes place in the site-permutation symmetric subspace of $\tilde{\mathcal{H}}=\otimes_j\tilde{\mathcal{H}}_j$ with ${\rm dim}(\tilde{\mathcal{H}}_j)=4$.

\section{Bounds on the entanglement Page curves}
In this section, we provide further details on the Page curve bound for the case of the von Neumann entanglement entropy
\begin{equation}\label{eq:bound_von_Neumann_SM}
S_{1}(t)\leq {\rm min}\left[|A|, S^{\rm PE}_1(|\psi\rangle),{|A|} \sqrt{1-z^2}+O(1)\right]\,.
\end{equation}
We derive in particular the last term, as the first two terms in the right-hand side follow immediately from the case of R\'enyi entropies derived in the main text.

First, setting $\delta=\sqrt{1-z^2}$, where $z=|\braket{\psi_0|a_N}|$, we see that $\delta$ is equal to the trace distance $\Delta(\ket{a_N}\bra{a_N},\ket{\psi_0}\bra{\psi_0})$ between $\ket{a_N}\bra{a_N}$ and $\ket{\psi_0}\bra{\psi_0}$. Therefore, using the contractivity of the trace distance with respect to the partial trace, we have
\begin{equation}
   \delta_A\equiv \Delta({\rm Tr}_{\bar{A}}[|a_N\rangle\langle a_N|],\rho_A= {\rm Tr}_{\bar{A}}[|\psi_0\rangle\langle \psi_0|)\leq \delta\,.
\end{equation}
Now, ${\rm Tr}_{\bar{A}}[|a_N\rangle\langle a_N|=(\ket{+}\bra{+})^{\otimes |A|}$ is a pure state. Thus, using the Fannes inequality~\cite{audenaert2007sharp}, we get
\begin{equation}
 S_1(\rho_A)\leq |A|\delta +O(1) \,,
\end{equation}
where we used that the binary entropy is $H(\delta_A,1-\delta_A)$ is in $[0,1]$ and $\delta_A\leq \delta$. Therefore, assuming $|A|\leq L/2$, we obtain
\begin{equation}
    S_1(\rho_A)\leq  |A| \sqrt{1-z^2}+O(1)\,.
\end{equation}

\section{Derivation of the system of differential equations}

Here, we show that the averaged purity can be obtained as the solution to a system of differential equations. We will start with the circuit of random permutation and random-phase gates, as it is technically simpler.

\subsection{Random Permutation Circuit with Random Phases (RPPC)}
\label{sec:Random Permutation Circuit with Random Phases (RPPC)}

In this section, we derive a system of differential equations that describe the time evolution of the averaged purity $\mathcal{P}_A(t)$:
\begin{equation}
\label{eq:purity_swap_state}
\mathcal{P}_A(t):=\text{Tr}_A\mathbb{E}[\rho_A^2(t)]=\braket{X_A|\mathbb{E}\left[\rho(t)\otimes\rho(t)\right]}\,,
\end{equation}
 for a 2-local random circuit consisting of random permutations with the addition of random phases. By choosing a site-permutation invariant initial state, the purity at time $t=0$ depends only on $n = |A|$, $\mathcal{P}_A(0)=\mathcal{P}_n(0)$. Because the Lindbladian \eqref{eq:Lindbladian_continuum_def_SM} is site-permutation invariant, the evolved state will also retain this property.

The presence of additional random phases reduces the dimension of the invariant subspace of $\mathcal{U}_{i,k}$, which is now spanned by the states
\begin{equation}
\ket{I^{+}I^{+}}_{j,k}\,,\quad\ket{I^{-}I^{-}}_{j,k}\,,\quad\ket{I^{0}I^{0}}_{j,k}\,,  
\end{equation}
where
\begin{equation}
    \ket{I^{\alpha}I^{\alpha}}_{j,k} = \ket{I˘^{\alpha}}_j \ket{I˘^{\alpha}}_k\,,\qquad \alpha = +, - , 0\,.
\end{equation}
Because of site-permutation invariance, in order to obtain a closed system of ODEs, we only need to introduce the $\binom{N+2}{2}$ states:
\begin{equation}\label{eq:G_state_RPP_original}
\ket{G_{n_-,n_+,n_0}} := \frac{1}{N!} \sum_{\pi \in \mathcal{S}_N} \pi \ket{I^-}^{\otimes n_-} \otimes \ket{I^+}^{\otimes n_+} \otimes \ket{I^0}^{\otimes n_0}\,,
\end{equation}
where we employ the same symbol $\pi$ to denote a permutation and its unitary implementation on a state. We then define the functions
\begin{equation}
\label{eq:G_function_def}
G_{n_-,n_+,n_0}(t) = \braket{G_{n_-,n_+,n_0}|\rho(t)\otimes \rho(t)} ,
\end{equation}
where $n_-$, $n_+$, $n_0$ are non-negative integers such that $n_- + n_+ + n_0=N$. The purity is recovered by setting $n_0=0$: 
\begin{equation}
\mathcal{P}_n(t)= \braket{G_{n,N-n,0}|\mathbb{E}\left[\rho(t)\otimes\rho(t)\right]}\, = \braket{X_n|\mathbb{E}\left[\rho(t)\otimes\rho(t)\right]},
\end{equation}
where $\ket{X_n} = (N!)^{-1} \sum_{\pi \in \mathcal{S}_N} \pi \ket{I^-}^{\otimes n}\ket{I^+}^{\otimes(N-n)}$ is the symmetrized swap state. The system of differential equations is:
\begin{equation}
\label{eq:purity_ODE_system}
\begin{split}
 \frac{d G_{n_-,n_+,n_0}(t)}{dt} =& \braket{G_{n_-,n_+,n_0}(t)|(-\mathcal{L})\mathbb{E}\left[\rho(t)\otimes\rho(t)\right]} \\
 =& \frac{\lambda[(n_-)^2+(n_+)^2+(n_0)^2-N^2]}{N-1}G_{n_-,n_+,n_0}(t) \\
 +&\frac{2\lambda n_-n_+}{(N-1)(q+1)}\left[G_{n_--1,n_++1,n_0}(t)+G_{n_-+1,n_+-1,n_0}(t)+(q-1)G_{n_--1,n_+-1,n_0+2}(t)\right] \\
 +&\frac{2\lambda n_0n_-}{(N-1)(q+1)}(G_{n_-+1,n_+,n_0-1}(t)+q\,G_{n_--1,n_+,n_0+1}(t)) \\ +&\frac{2\lambda n_0n_+}{(N-1)(q+1)}(G_{n_-,n_++1,n_0-1}(t)+q\,G_{n_-,n_+-1,n_0+1}(t)).
\end{split}
\end{equation}
The system closes by adopting the convention that $G_{n_-,n_+,n_0}(t)=0$ whenever any of the three indices is $-1$ or $N+1$. In particular, the equation for the purity is:
\begin{equation}
    \frac{d\mathcal{P}_n(t)}{dt}= 
    -\frac{2\lambda n(N-n)}{N-1}\mathcal{P}_n(t) 
    +\frac{2\lambda n (N-n)}{(N-1)(q+1)}
    \left(\mathcal{P}_{n-1}(t) + \mathcal{P}_{n+1}(t)\right) 
    + \frac{2\lambda n(N-n)(q-1)}{(N-1)(q+1)}G_{n-1,N-n-1,2}(t).
\end{equation}

To prove~\eqref{eq:purity_ODE_system} we rewrite Eq.~\eqref{eq:G_state_RPP_original} as
\begin{equation}
\label{eq:G_state_RPP_permutation_explicit}
\ket{G_{n_-,n_+,n_0}} = \frac{1}{N!} \sum_{\pi \in \mathcal{S}_N} 
  \bigotimes_{i \in A_\pi^-}\ket{I^-}_i 
  \bigotimes_{i \in A_\pi^+}\ket{I^+}_i \bigotimes_{i \in A_\pi^0}\ket{I^0}_i\,,
\end{equation}
where $A_\pi^- =\{\pi(1),\dots,\pi(n_-)\}$, $A_\pi^+ =\{\pi(n_-+1),\dots,\pi(n_-+n_+)\}$, $A_\pi^0 =\{\pi(n_-+n_++1),\dots,\pi(N)\}$, and we note that, for a fixed permutation $\pi$, there are exactly $n_-n_+$ ways one could remove one index from the set $A_{\pi}^{-}$ and one from the set $A_{\pi}^{+}$, and move them to $A_{\pi}^{0}$. This gives the following representation of $\ket{G_{n_--1,n_+-1,n_0+2}}$ in terms of the r.h.s.\ of Eq.~\eqref{eq:G_state_RPP_permutation_explicit},
\begin{equation}
\ket{G_{n_--1,n_+-1,n_0+2}} = \frac{1}{N!} \sum_{\pi \in \mathcal{S}_N} 
  \frac{1}{n_-n_+}\sum_{\substack{j \in A_\pi^- \\ 
  k \in A_\pi^+}}\,\bigotimes_{\substack{i \in A_\pi^- \\ i \ne j}}\ket{I^-}_i 
  \bigotimes_{\substack{i \in A_\pi^+ \\ i \ne k}}\ket{I^+}_i 
  \bigotimes_{i \in A_\pi^0 \cup \{j,k\}}\ket{I^0}_i\,,
\end{equation}
and the rest of the states appearing in Eq.~\eqref{eq:purity_ODE_system} can be expressed in an analogous form. Thus, from:
\begin{equation}
  \begin{aligned}
    \mathcal{U}_{j,k}\ket{I^-I^-}_{j,k} &= \ket{I^-I^-}_{j,k}\,, \\
    \mathcal{U}_{j,k}\ket{I^+I^+}_{j,k} &= \ket{I^+I^+}_{j,k}\,, \\
    \mathcal{U}_{j,k}\ket{I^0I^0}_{j,k} &= \ket{I^0I^0}_{j,k}\,, \\
    \mathcal{U}_{j,k}\ket{I^-I^0}_{j,k} &= \frac{1}{q+1}(\ket{I^-I^-}_{j,k}+q\,\ket{I^0I^0}_{j,k})\,, \\
    \mathcal{U}_{j,k}\ket{I^+I^0}_{j,k} &= \frac{1}{q+1}(\ket{I^+I^+}_{j,k}+q\,\ket{I^0I^0}_{j,k})\,, \\
    \mathcal{U}_{j,k}\ket{I^+I^-}_{j,k} &= \frac{1}{q+1}[\ket{I^-I^-}_{j,k}+\ket{I^+I^+}_{j,k}+(q-1)\,\ket{I^0I^0}_{j,k}]\,,
  \end{aligned}
\end{equation}
we obtain:
\begin{equation}
  \begin{split}
    &\sum_{1\le j < k \le N}\bra{G_{n_-,n_+,n_0}(t)}\mathcal{U}_{j,k} \\ 
    &=\frac{1}{N!}\sum_{\pi \in \mathcal{S}_N} 
    \left(
    \sum_{\substack{j,k\in A_\pi^- \\ j < k}} 
    + \sum_{\substack{j,k\in A_\pi^+ \\ j < k}} 
    + \sum_{\substack{j,k\in A_\pi^0 \\ j < k}}
    + \sum_{\substack{j \in A_\pi^- \\ k \in A_\pi^+}} 
    + \sum_{\substack{j \in A_\pi^- \\ k \in A_\pi^0}} 
    + \sum_{\substack{j \in A_\pi^+ \\ k \in A_\pi^0}} 
    \right)
    \bigotimes_{i \in A_\pi^-}\bra{I^-}_i \bigotimes_{i \in A_\pi^+}\bra{I^+}_i 
    \bigotimes_{i \in A_\pi^0}\bra{I^0}_i \,\mathcal{U}_{j,k}\\
    &= \left(\frac{n_{-}(n_{-}-1)}{2}
    +\frac{n_{+}(n_{+}-1)}{2}
    +\frac{n_{0}(n_{0}-1)}{2}\right)
    \bra{G_{n_-,n_+,n_0}(t)} \\ 
    &+ \frac{n_-n_+}{q+1}
    \left(\bra{G_{n_--1,n_++1,n_0}(t)} 
    + \bra{G_{n_-+1,n_+-1,n_0}(t)}+(q-1)\bra{G_{n_--1,n_+-1,n_0+2} (t)}\right) \\
    &+\frac{n_-n_0}{q+1}
    \left(\bra{G_{n_-+1,n_+,n_0-1}(t)} + q\,\bra{G_{n_{-}-1,n_{+},n_0+1}}\right)
    +\frac{n_+n_0}{q+1}
    \left(\bra{G_{n_-,n_++1,n_0-1}(t)} + q\,\bra{G_{n_-,n_+-1,n_0+1}}\right).
  \end{split}
\end{equation}
Equation~\eqref{eq:purity_ODE_system} then follows straightforwardly from the definition~\eqref{eq:Lindbladian_continuum_def_SM}. 

Let us now derive the initial conditions for the homogeneous product state:  
\begin{equation} \label{eq:homogeneous_product_state}
  \ket{\Psi_0} = \left(\sum_{a = 0}^{q-1}\lambda_a\ket{a}\right)^{\otimes N},
  \qquad \sum_{a=0}^{q-1} |\lambda_a|^2  = 1,
\end{equation}
which corresponds to:
\begin{equation} \label{eq:CJ_product_state}
  \ket{\rho_0\otimes\rho_0} = \bigotimes_{j = 1}^{N}
    \left[\sum_{a,b,c,d = 0}^{q-1}\lambda_a^* \lambda_b \lambda_c^* \lambda_d\ket{abcd}_j\right].
\end{equation}

The computation is straightforward:
\begin{equation}
\begin{split}
&G_{n_-,n_+,n_0}(0) =  \braket{G_{n_-,n_+,n_0}|\rho_0\otimes \rho_0} \\
=& \frac{1}{N!} \sum_{\pi \in S_N} \left(\bra{I^-}\sum_{a,b,c,d}\lambda_a^* \lambda_b \lambda_c^* \lambda_d\ket{abcd}\right)^{n_-}\left(\bra{I^+}\sum_{a,b,c,d}\lambda_a^* \lambda_b \lambda_c^* \lambda_d\ket{abcd}\right)^{n_+}\left(\bra{I^0}\sum_{a,b,c,d}\lambda_a^* \lambda_b \lambda_c^* \lambda_d\ket{abcd}\right)^{n_0} \\
=&\frac{1}{N!}\sum_{\pi \in \mathcal{S}_N} \left(\sum_{a=0}^{q-1}|\lambda_a|^4\right)^{n_0} = \left(\sum_{a=0}^{q-1}|\lambda_a|^4\right)^{n_0},
\end{split}
\end{equation}
a result consistent with $\mathcal{P}_n(0)=1$ for this state. In particular, for a qubit state:
\begin{equation}
\ket{\Psi_0}=(\cos\theta \ket{0} + \sin\theta \ket{1})^{\otimes N}\,,
\end{equation}
\begin{equation}\label{eq:qubit-init-vals}
G_{n_-,n_+,n_0}(0)=\left(1-\frac{1}{2}\sin^2(2\theta)\right)^{n_0},
\end{equation}
so that $G_{n_-,n_+,n_0}(0)=1$ for $\theta=0, \pi/2$ (corresponding to \textit{classical states} $\ket{\Psi_0} = \ket{a}^{\otimes N}$) and the minimum of the initial value is achieved when $\theta=\pi/4$, for which $G_{n_-,n_+,n_0}(0)=2^{-n_0}$. In the $q$-dimensional case, one has:
\begin{equation}
\label{eq:G_func_homogeneous_tilted_state}
G_{n_-,n_+,n_0}(0)=q^{-n_0}\, \quad \text{for} \quad \ket{\Psi_0}=\left(\sum_{a=0}^{q-1}\frac{e^{i\theta_a}}{\sqrt{q}}\ket{a}\right)^{\otimes N}.
\end{equation}

\subsection{Entangled Initial States and Anticoncentration}
\label{sec: Initial states}

In this section, we compute the initial conditions for a 2-local random permutation circuit with random phases, that is, the initial conditions for the system \eqref{eq:purity_ODE_system}, when the initial state is a site-permutation invariant entangled state. In particular, we focus on the following:
\begin{enumerate}
    \item The GHZ state: 
    \begin{equation}
    \label{eq:GHZ_state_def}
    \ket{\text{GHZ}} = \frac{1}{\sqrt{q}}\sum_{a=0}^{q-1}\ket{a}^{\otimes N};
    \end{equation}
    \item The Dicke states ($q=2$)
    \begin{equation}
    \label{eq:Dicke_state_def}
    \ket{D} = \sum_{k=0}^N c_k \ket{D_k},\quad  \ket{D_k} = \binom{N}{k}^{-1/2} \sum_{\pi} \pi \ket{0}^{\otimes(N-k)}\otimes \ket{1}^{\otimes k}, \quad \sum_{k=0}^N |c_k|^2=1\,,
    \end{equation}
    where the sum is over the distinct permutations of the qubits. Notice that by considering $\ket{0}$, $\ket{1}$ as two spin states of a spin-1/2 chain, the state $\ket{W} \equiv \ket{D_1}$ is a one-magnon state with zero momentum. Its purity in a bipartiton, as well as that of $\ket{D_2}$, was computed in \cite{castro2018entanglement}.
\end{enumerate}  

When the initial state is not a product state, a direct use of the Choi-Jamiolkowski isomorphism is generally not the most efficient way to proceed. This is because the expression of the two-replica state $\ket{\rho\otimes \rho}$ may be quite cumbersome. Instead, we will derive an alternative expression for the functions $G_{n_-,n_+,n_0}$ which is useful in the case of permutation-invariant entangled states. Let us define a tripartite Hilbert space:
\begin{equation}
\label{eq:tripartite_Hilbert_space}
\mathcal{H}=\mathcal{H}_-\otimes\mathcal{H}_+\otimes\mathcal{H}_0 := \left(\mathcal{H}_\text{loc}\right)^{\otimes n_-}\otimes\left(\mathcal{H}_\text{loc}\right)^{\otimes n_+}\otimes\left(\mathcal{H}_\text{loc}\right)^{\otimes n_0}, \quad \mathcal{H}_\text{loc}\simeq \mathbb{C}^q, \quad n_- + n_+ + n_0 = N,
\end{equation}
corresponding to three non overlapping regions $A_-$, $A_+$, $A_0$, of sizes $|A_-|=n_-$, $|A_+|=n_+$, $|A_0|=n_0$. We then consider the functions $G_{n_-,n_+,n_0}(t)$ defined in \eqref{eq:G_function_def}, and let $\rho(t)$ be a permutation-invariant state. By using the explicit expression \eqref{eq:Choi_representation_SM} for the 2-replica state $\ket{\rho(t)\otimes\rho(t)}$, after some algebra we arrive at:
\begin{equation}
\label{eq:G_functions_IPR_representation}
G_{n_-,n_+,n_0}(t) = \sum_{\{x_i\}_{i \in A_0}} \text{Tr}_{\mathcal{H}_-}(\bra{\{x_i\}}\text{Tr}_{\mathcal{H}_{+}}\rho(t)\ket{\{x_i\}})^2,
\end{equation}
where the states $\ket{\{x_i\}}$, for $x_i=0,\dots,q-1$ and $i \in A_0$,  form the computational basis of $\mathcal{H}_0$. 
For $n_-=n$, $n_+=N-n$, $n_0=0$, the right-hand side of equation \eqref{eq:G_functions_IPR_representation} yields the averaged purity $\mathcal{P}_n(t)$, as expected. On the other hand, for $\rho(t)=\ket{\psi(t)}\bra{\psi(t)}$:
\begin{equation}
  G_{0,0,N}(t)= \sum_{\{x_i\}_{i=1,\dots,N}} (\bra{\{x_i\}}\rho(t)\ket{\{x_i\}})^2 = \sum_{\{x_i\}} |\braket{\{x_i\}|\psi(t)}|^4=I_2(\ket{\psi(t)}),
\end{equation}
where $I_2(\ket{\psi})$ is the second inverse participation ratio of the state $\ket{\psi}$ (cf.\ Eq.\ \eqref{eq:inverse_participation}). For generic values of $n_-$, $n_+$, $n_0$ an intermediate situation arises, whereby partial traces are taken over the regions $A_-$, $A_+$ while the off-diagonal matrix elements in the region $A_0$ are washed away. We observe that this is consistent with the result \eqref{eq:G_func_homogeneous_tilted_state} for maximally delocalized product states, for which it is known that $I_2 = q^{-N}$. 

By using equation \eqref{eq:G_functions_IPR_representation}, the computation of the functions $G_{n_-,n_+,n_0}$ for $\ket{\text{GHZ}}$ is straightforward. Indeed, in this case $\rho=\ket{\text{GHZ}}\!\bra{\text{GHZ}}$ and
\begin{equation}
  \begin{gathered}
    \text{Tr}_{\mathcal{H}_+}\,\rho 
    = \frac{1}{q} \sum_{a,b=0}^{q-1} \text{Tr}_{\mathcal{H}_+} (\ket{a}\bra{b})^{\otimes N} 
    = \frac{1}{q}\sum_{a=0}^{q-1}(\ket{a}\bra{b})^{\otimes (N-n_+)},  \\
    \bra{\{x_i\}}\text{Tr}_{\mathcal{H}_{+}}\,\rho\ket{\{x_i\}} 
    = \frac{1}{q} 
    \sum_{a=0}^{q-1}(\ket{a}\bra{a})^{\otimes n_-}\prod_{i \in A_0} \delta_{x_i,a}\,.
  \end{gathered}
\end{equation}
Hence:
\begin{equation}
G_{n_-,n_+,n_0}(\ket{\text{GHZ}}) = \frac{1}{q^2} \sum_{a=0}^{q-1}\sum_{\{x_i\}_{i \in A_0}} \prod_{i \in A_0} \delta_{x_i, a} = \frac{1}{q}, 
\end{equation}
for all triples $(n_-,n_+,n_0)$.

Let us now consider the Dicke state \eqref{eq:Dicke_state_def}. The computation of $I_2(\ket{D})$ is straightforward, and yields:
\begin{equation}
I_2(\ket{D}) = \sum_{m=1}^N |c_m|^4 \binom{N}{m}^{-1}.
\end{equation}
This implies in particular that we can express the second PE of $\ket{D}$ as:
\begin{equation}
S_2^\text{PE}(\ket{D})=-\log I_2(\ket{D}) = \Delta_2 \log(2^N), \quad \Delta_2 = -\frac{1}{N}\log\left(\sum_{m=1}^N |c_m|^4\binom{N}{m}^{-1}\right).
\end{equation}
The quantity $\Delta_2$ is a fractal dimension $0 \leq \Delta_2 \leq1$, which signals the amount of delocalization (or anticoncentration) of the state: the latter is maximally localized when $\Delta_2=0$ and maximally delocalized when $\Delta_2=1$.

Computing the initial conditions $G_{n_-,n_+,n_0}(0)$ for the state $\ket{D}$ when $(n_-,n_+,n_0)$ is a generic triplet is a difficult task. We limit ourselves to the case of $\ket{D}=c_1\ket{D_1} + c_2\ket{D_2}$, $|c_1|^2 + |c_2|^2=1$, that is a linear superposition of a one- and a two-magnon state. First, we observe that by defining for any region $A \in \{A_-,A_+,A_0\}$:
\begin{equation}
\ket{D_k}_A = \binom{|A|}{k}^{-1/2} \sum_{\pi} \pi\ket{0}^{\otimes(|A|-k)}\otimes \ket{1}^{\otimes k}\, \in \mathcal{H}_\text{loc}^{\otimes |A|},
\end{equation}
then in the Hilbert space \eqref{eq:tripartite_Hilbert_space} we can write, for $k=0,\dots,N$:
\begin{equation}
\label{eq:Dicke_state_multipartite}
\ket{D_k} = \binom{N}{k}^{-\frac{1}{2}}\sum_{k_-=0}^k\sum_{k_+=0}^{k-k_-}\binom{n_-}{k_-}^{\frac{1}{2}}\binom{n_+}{k_+}^{\frac{1}{2}}\binom{n_0}{k_0}^{\frac{1}{2}}\ket{k_-,k_+,k_0}, \quad \ket{k_-,k_+,k_0} := \ket{D_{k_-}}_{A_-} \ket{D_{k_+}}_{A_+} \ket{D_{k_0}}_{A_0}\,,
\end{equation}
where $n_0 = N-n_--n_+$, and $k_0 = k-k_--k_+$ for every $k_-$, $k_+$. We also adopted the convention that $\binom{N}{k}=0$ if $k < 0$ or $k > N$. Thus, defininig the ratios:
\begin{equation}
r_-:=\frac{n_-}{N}, \quad r_+:=\frac{n_+}{N}, \quad r_0:=\frac{n_0}{N},
\end{equation}
we have:
\begin{align}
\ket{D_1} &= r_-^\frac{1}{2}\ket{1,0,0} +  r_+^\frac{1}{2}\ket{0,1,0} + r_0^\frac{1}{2}\ket{0,0,1}, \\
\ket{D_2}  &= \left[r_-\left(\frac{n_--1}{N-1}\right)\right]^{\frac{1}{2}} \ket{2,0,0} + \left[r_+\left(\frac{n_+-1}{N-1}\right)\right]^{\frac{1}{2}} \ket{0,2,0} + \left[r_0\left(\frac{n_0-1}{N-1}\right)\right]^{\frac{1}{2}} \ket{0,0,2} \nonumber \\
&+ \left(\frac{2 r_-n_+}{N-1}\right)^\frac{1}{2}\ket{1,1,0} + \left(\frac{2 r_-n_0}{N-1}\right)^\frac{1}{2}\ket{1,0,1} + \left(\frac{2 r_+n_0}{N-1}\right)^\frac{1}{2}\ket{0,1,1}.
\end{align}
We note that the expression \eqref{eq:Dicke_state_multipartite} can be generalized to compute multipartite entanglement, and it was first employed for the description of kink states in the ferromagnetic Ising chain in \cite{capizzi2025entanglement}.
The reduced density matrix of $\ket{D}$ in $\mathcal{H}_-\otimes \mathcal{H}_0$ is obtained as
\begin{equation}
\mkern-30mu
\begin{aligned}
\text{Tr}_{\mathcal{H}_+}\ket{D}\!\bra{D} &= 
  |c_1|^2(r_-\ket{1,0}\!\bra{1,0} + r_+\ket{0,0}\!\bra{0,0}+r_0\ket{0,1}\!\bra{0,1} ) \\ 
  &+ |c_2|^2 \left[r_-\left(\frac{n_--1}{N-1}\right)\ket{2,0}\!\bra{2,0}
  +r_-\left(\frac{n_+-1}{N-1}\right)\ket{0,0}\!\bra{0,0} 
  +r_0\left(\frac{n_0-1}{N-1}\right)\ket{0,2}\!\bra{0,2}\right. \\
  & \left. + \frac{2r_-n_+}{N-1}\ket{1,0}\!\bra{1,0} + \frac{2r_-n_+}{N-1}\ket{1,0}\!\bra{1,0} 
  + \frac{2r_-n_0}{N-1}\ket{1,1}\!\bra{1,1}\right] \\
  &+ c_1c_2^* \left[r_-\left(\frac{n_--1}{N-1}\right)^\frac{1}{2}\ket{1,0}\!\bra{2,0} 
  + r_+^\frac{1}{2}\left(\frac{2r_-n_+}{N-1}\right)\ket{0,0}\!\bra{1,0} 
  + r_0^\frac{1}{2}\left(\frac{2r_-n_0}{N-1}\right)\ket{0,1}\!\bra{1,0}\right] + \text{h.c.} \\
  &+ \text{off-diagonal in } \mathcal{H}_0\,.
\end{aligned}
\mkern-30mu
\end{equation}
In the above expression, we only list those terms with non-zero diagonal matrix elements in $\mathcal{H}_0$. In order to evaluate the diagonal matrix elements in $\mathcal{H}_0$ we make use of the following decomposition of the Hilbert space,
\begin{equation}
\mathcal{H}_0 = \oplus_{k=0}^{n_0} \mathcal{H}_0^{(k)}, 
  \qquad \mathcal{H}_0^{(k)} 
  = \text{Span}\left\{\ket{\{x_i\}}_{i\in A_0}\,\middle\rvert \,x_i \in \{0,1\}\,, 
  \sum_{i \in A_0} x_i =k\right\}.
\end{equation}
The space $\mathcal{H}_0^{(k)}$ is the sector of $\mathcal{H}_0$ in which the number of local states $\ket{1}$ in the region $A_0$ is fixed to $k$. As expected, $\sum_{k=0}^{n_0} \text{dim} \mathcal{H}_0^{(k)}=\sum_{k=0}^{n_0} \binom{n_0}{k} = 2^{n_0}$. By using this above decomposition, for $\ket{\{x_i\}},\,\ket{D_{k_0}}\,\in \mathcal{H}_0$:
\begin{equation}
\braket{\{x_i\}|D_{k_0}} = \binom{n_0}{k_0}^{-\frac{1}{2}}\delta_{\sum_{i}x_i-k_0,0}\,.
\end{equation}
Therefore, it is a matter of algebra to obtain from \eqref{eq:G_functions_IPR_representation}:
\begin{equation}
  \mkern-20mu
  \begin{aligned}
    G_{n_-,n_+,n_0}(c_1\ket{D_1}+c_2\ket{D_2})
    =&|c_1|^4\left(r_+^2+r_-^2 + \frac{r_0^2}{N}\right)
    +|c_2|^4\left[r_+^2\left(\frac{n_+-1}{N-1}\right)^2+r_-^2\left(\frac{n_--1}{N-1}\right)^2
    +\left(\frac{2r_-n_+}{N-1}\right)^2\right. \\
    +& \left. \frac{1}{n_0}\left(\frac{2r_+n_0}{N-1}\right)^2 +\frac{1}{n_0}\left(\frac{2r_-n_0}{N-1}\right)^2 + \frac{2}{n_0(n_0-1)}\left(r_0\left(\frac{n_0-1}{N-1}\right)\right)^2\right] \\
    +&2|c_1|^2|c_2|^2\left[r_+^2\left(\frac{n_+-1}{N-1}\right)+r_-^2\left(\frac{n_--1}{N-1}\right)+ \frac{2r_-^2n_+}{N-1} + \frac{2r_+^2n_-}{N-1} + \frac{2r_+r_0}{N-1} + \frac{2r_-r_0}{N-1}\right]. 
  \end{aligned}
  \mkern-50mu
\end{equation}

\subsection{Random Permutation Circuit with Initial Random Product States}
 We now derive the system of differential equations for the averaged purity in a 2-local Random Permutation Circuit with an initial random product state, that is, a state in which local (on-site) scrambling is applied before the action of the 2-local gates. We focus on qubits ($q=2$). Since permutation gates acting on two-qubit states are Clifford, the considerations of Section \ref{sec:local dimensional reduction} apply, and the evolution takes place in an effective 4-dimensional Hilbert space.

\subsubsection{Differential equations}

By inspection, one can show that the effective local Hilbert space is spanned the following states in the four-replica space
\begin{equation}
  \begin{aligned}
    \ket{J^0}&=\frac{1}{\sqrt{2}}(\ket{0000}+\ket{1111})\,,&\qquad
    \ket{J^+}&=\frac{1}{\sqrt{2}}(\ket{0011}+\ket{1100})\,,\\
    \ket{J^x}&=\frac{1}{\sqrt{2}}(\ket{0101}+\ket{1010})\,&
    \ket{J^-}&=\frac{1}{\sqrt{2}}(\ket{1001}+\ket{0110})\,.
  \end{aligned}
\end{equation}
Indeed, defining
\begin{equation}
  \begin{aligned}
    \ket{v_0}_{i,j}&=\ket{J^0}_i\ket{J^0}_j\,\\
    \ket{v_1}_{i,j}&=\frac{1}{\sqrt{3}}(\ket{J^0}_i\ket{J^+}_j+\ket{J^+}_i\ket{J^0}_j+\ket{J^+}_i\ket{J^+}_j)\,,\\
    \ket{v_2}_{i,j}&=\frac{1}{\sqrt{3}}(\ket{J^0}_i\ket{J^x}_j+\ket{J^x}_i\ket{J^0}_j+\ket{J^x}_i\ket{J^x}_j)\,,\\
    \ket{v_3}_{i,j}&=\frac{1}{\sqrt{3}}(\ket{J^0}_i\ket{J^-}_j+\ket{J^-}_i\ket{J^0}_j+\ket{J^-}_i\ket{J^-}_j)\,,\\
    \ket{v_4}_{i,j}&=\frac{1}{\sqrt{6}}(\ket{J^+}_i \ket{J^x}_j + \ket{J^x}_i\ket{J^+}_j+\ket{J^+}_i\ket{J^-}_j+\ket{J^-}_i\ket{J^+}_j +
    \ket{J^x}_i\ket{J^-}_j +\ket{J^-}_i\ket{J^x}_j)\,,
  \end{aligned}
\end{equation}
one can show
\begin{equation}\label{eq:Ujk-rpc-typical-sm}
\tilde{\mathcal{U}}_{i,j} = \Pi_i\Pi_j \,\mathcal{U}_{i,j}\,\Pi_i\Pi_j
  =\sum_{k=0}^4 \ket{v_{k}}_{i,j}\bra{v_k}_{i,j},
\end{equation}
with the projectors defined in~\eqref{eq:projector_Clifford_SM}. Then, the Lindbladian $\tilde{\mathcal{L}}$ is simply obtained from \eqref{eq:Lindbladian_continuum_def_SM} by replacing the averaged gate $\mathcal{U}_{i,j}$ with $\tilde{\mathcal{U}}_{i,j}$.

The time evolution of the averaged purity in a 2-local random permutation circuit with a random initial state is obtained by solving a system of differential equations for the following $\binom{N+3}{3}$ functions
\begin{equation}
  G_{p,q,r,s}(t):=
  \braket{G_{p,q,r,s}|\mathbb{E}[\rho(t)\otimes\rho(t)]},
  \qquad p,q,r,s \in \mathbb{N}_0,\quad p+q+r+s=N,
\end{equation}
where:
\begin{equation}
\ket{G_{p,q,r,s}} := \frac{1}{N!} \sum_{\pi \in \mathcal{S}_N} \pi \ket{I^-}^{\otimes p} \otimes \ket{I^+}^{\otimes q} \otimes \ket{I^0}^{\otimes r}\otimes \ket{I^x}^{\otimes s}\,,
\end{equation}
and we introduced the state $\ket{I^x}=\sum_{a,b=0}^1\ket{abab}$.
The $\ket{I^\alpha}$ and $\ket{J^\beta}$ states are related via:
\begin{equation}\label{eq:DefIalphaJalphaRP}
\ket{J^0}=\frac{\ket{I^0}}{\sqrt{2}}, \quad \ket{J^+}=\frac{\ket{I^+}-\ket{I^0}}{\sqrt{2}}, \quad \ket{J^x}=\frac{\ket{I^x}-\ket{I^0}}{\sqrt{2}}, \quad \ket{J^-}=\frac{\ket{I^-}-\ket{I^0}}{\sqrt{2}}.
\end{equation}

A closed system of ODEs can be obtained by imposing the condition that $G_{p,q,r,s}(t)=0$ whenever any of the indices is either $0$ or $N+1$. In order to derive the equations, we first compute the action of $\tilde{\mathcal{U}}_{i,j}$ on the replica states $\ket{I^\alpha I^\beta}_{i,j}$, observing that $\tilde{\mathcal{U}}_{i,j} \ket{I^\alpha I^\beta}_{i,j} = \tilde{\mathcal{U}}_{i,j} \ket{I^\beta I^\alpha}_{i,j}$: 
\begin{equation}
  \begin{aligned}
    \tilde{\mathcal{U}}_{i,j}\ket{I^-I^-}_{i,j} &= \ket{I^-I^-}_{i,j}\,, \\
    \tilde{\mathcal{U}}_{i,j}\ket{I^+I^+}_{i,j} &= \ket{I^+I^+}_{i,j}\,, \\
    \tilde{\mathcal{U}}_{i,j}\ket{I^xI^x}_{i,j} &= \ket{I^xI^x}_{i,j}\,, \\
    \tilde{\mathcal{U}}_{i,j}\ket{I^0I^0}_{i,j} &= \ket{I^0I^0}_{i,j}\,, \\
    \tilde{\mathcal{U}}_{i,j}\ket{I^-I^0}_{i,j} &= \ket{I^0I^0}_{i,j} 
    + \frac{1}{3}(\ket{I^-I^-}_{i,j}-\ket{I^0I^0}_{i,j})\,, \\
    \tilde{\mathcal{U}}_{i,j}\ket{I^+I^0}_{i,j} &= \ket{I^0I^0}_{i,j} 
    + \frac{1}{3}(\ket{I^+I^+}_{i,j}-\ket{I^0I^0}_{i,j})\,, \\
    \tilde{\mathcal{U}}_{i,j}\ket{I^xI^0}_{i,j} &= \ket{I^0I^0}_{i,j} 
    + \frac{1}{3}(\ket{I^xI^x}_{i,j}-\ket{I^0I^0}_{i,j})\,, \\
    \tilde{\mathcal{U}}_{i,j}\ket{I^-I^+}_{i,j} &= \frac{1}{3}(\ket{I^-I^-}_{i,j}
    +\ket{I^+I^+}_{i,j} +\ket{I^0I^0}_{i,j}) + \frac{2}{\sqrt{6}} \ket{v_4}_{i,j}\,, \\
    \tilde{\mathcal{U}}_{i,j}\ket{I^-I^x}_{i,j} &= \frac{1}{3}(\ket{I^-I^-}_{i,j}
    +\ket{I^xI^x}_{i,j} +\ket{I^0I^0}_{i,j}) + \frac{2}{\sqrt{6}} \ket{v_4}_{i,j}\,, \\
    \tilde{\mathcal{U}}_{i,j}\ket{I^+I^x}_{i,j} &= \frac{1}{3}(\ket{I^+I^+}_{i,j}
    +\ket{I^xI^x}_{i,j} +\ket{I^0I^0}_{i,j}) + \frac{2}{\sqrt{6}} \ket{v_4}_{i,j}\,.
  \end{aligned}
\end{equation}
Then, proceeding as in Section \ref{sec:Random Permutation Circuit with Random Phases (RPPC)}, we rewrite $\ket{G_{p,q,r,s}}$ as
\begin{equation}
  \ket{G_{p,q,r,s}} = \frac{1}{N!} \sum_{\pi \in \mathcal{S}_N} 
  \bigotimes_{i \in A_\pi^p}\ket{I^-}_i \bigotimes_{i \in A_\pi^q}\ket{I^+}_i 
  \bigotimes_{i \in A_\pi^r}\ket{I^0}_i\bigotimes_{i \in A_\pi^s}\ket{I^x}_i\,,
\end{equation}
where
\begin{equation}
  \begin{aligned}
    A_\pi^p &=\{\pi(1),\dots,\pi(p)\},  &\qquad
    A_\pi^q &=\{\pi(p+1),\dots,\pi(p+q)\}, \\ 
    A_\pi^r &=\{\pi(p+q+1),\dots,\pi(p+q+r)\}, &\qquad
    A_\pi^s &=\{\pi(p+q+r+1),\dots,\pi(N)\}.
  \end{aligned}
\end{equation}
This follows from appropriately splitting the sum over indices in:
\begin{equation}
  \begin{aligned}
    &\sum_{1\le i < j \le N}\bra{G_{p,q,r,s}(t)}
    \tilde{\mathcal{U}}_{i,j}\ket{\mathbb{E}[\rho(t)\otimes \rho(t)]} \\ 
    &=\frac{1}{N!}\sum_{\pi \in \mathcal{S}_N} \left(\sum_{\substack{i,j\in A_\pi^p \\ i < j}} 
    + \sum_{\substack{i,j\in A_\pi^q \\ i < j}} 
    + \sum_{\substack{i,j\in A_\pi^r \\ i < j}} 
    + \sum_{\substack{i,j\in A_\pi^s \\ i < j}} 
    + \sum_{\substack{i \in A_\pi^p \\ j \in A_\pi^q}}
    + \sum_{\substack{i \in A_\pi^p \\ j \in A_\pi^r}} 
    + \sum_{\substack{i \in A_\pi^p \\ j \in A_\pi^s}} 
    + \sum_{\substack{i \in A_\pi^q \\ j \in A_\pi^r}} 
    + \sum_{\substack{i \in A_\pi^q \\ j \in A_\pi^s}} 
    + \sum_{\substack{i \in A_\pi^r \\ j \in A_\pi^s}}\right) \\ 
    & \bigotimes_{k \in A_\pi^p}\bra{I^-}_k \bigotimes_{k \in A_\pi^q}\bra{I^+}_k 
    \bigotimes_{k \in A_\pi^r}\bra{I^0}_k\bigotimes_{k \in A_\pi^s}\bra{I^x}_k \,
    \tilde{\mathcal{U}}_{i,j}\, \ket{\mathbb{E}[\rho(t)\otimes \rho(t)]}\rangle.
  \end{aligned}
\end{equation}
After some algebra, the above yields the following system of equations:
\begin{equation}
  \begin{aligned}
    \label{eq:ODE_system_RPP_q=2}
    \frac{d G_{p,q,r,s}(t)}{dt} =& 
    \braket{G_{p,q,r,s}(t)|(-\tilde{\mathcal{L}})\mathbb{E}\left[\rho(t)\otimes\rho(t)\right]}\\
    =& -\lambda N G_{p,q,r,s}(t)+\frac{\lambda[p(p-1)+q(q-1)+r(r-1)+s(s-1)}{N-1}G_{p,q,r,s}(t)\\
    +&\frac{2\lambda p s}{3(N-1)}\left[G_{p+1,q,r,s-1}(t)+2G_{p-1,q,r,s+1}(t)\right]\\
    +&\frac{2\lambda q s}{3(N-1)}\left[G_{p,q+1,r,s-1}(t)+2G_{p,q-1,r,s+1}(t)\right]\\ 
    +&\frac{2\lambda r s}{3(N-1)}\left[G_{p,q,r+1,s-1}(t)+2G_{p,q,r-1,s+1}(t)\right]\\
    +&\frac{2\lambda p q}{3(N-1)}\left[G_{p+1,q-1,r,s}(t)+G_{p-1,q+1,r,s}(t)+G_{p-1,q,r+1,s}(t)
    +G_{p,q-1,r+1,s}(t) +G_{p,q,r,s}(t)\right.\\
    &\left. +4G_{p-1,q-1,r,s+2}(t)-2G_{p-1,q,r,s+1}(t)
    -2G_{p,q-1,r,s+1}(t)-2G_{p-1,q-1,r+1,s+1}(t)\right]\\
    +&\frac{2\lambda p r}{3(N-1)}\left[G_{p+1,q,r-1,s}(t)+G_{p-1,q,r+1,s}(t)
    +G_{p-1,q+1,r,s}(t)  +G_{p,q+1,r-1,s}(t) +G_{p,q,r,s}(t)\right.\\ 
    &\left. +4G_{p-1,q,r-1,s+2}(t)-2G_{p,q,r-1,s+1}(t)-2G_{p-1,q+1,r-1,s+1}(t)
    -2G_{p-1,q,r,s+1}(t)\right]\\
    +&\frac{2\lambda q r}{3(N-1)}\left[G_{p,q+1,r-1,s}(t)+G_{p,q-1,r+1,s}(t)
    +G_{p+1,q-1,r,s}(t)  +G_{p+1,q,r-1,s}(t) +G_{p,q,r,s}(t)\right.\\
    &\left. +4G_{p,q-1,r-1,s+2}(t)-2G_{p+1,q-1,r-1,s+1}(t)
    -2G_{p,q,r-1,s+1}(t)-2G_{p,q-1,r,s+1}(t)\right].
  \end{aligned}
\end{equation}

We consider a random initial state,
\begin{equation}
  \ket{\psi_0} = \bigotimes_{j=1}^N\left(U_j\ket{0}_j\right), \qquad U_j \in U(2),
\end{equation}
so that
\begin{equation}\label{eq:typical-init-state_sm}
  \begin{aligned}
    \ket{\mathbb{E}[\rho_0\otimes \rho_0]}
    &= \bigotimes_{j=1}^N \left[\mathbb{E}(U^*_j\otimes U_j\otimes U^*_j 
    \otimes U_j)\ket{0000}_j\right] \\
    &= \bigotimes_{j=1}^N \left[\frac{1}{d^2-1}
    \left(\ket{I^+}_{j\,j}\bra{I^+}+\ket{I^-}_{j\,j}\bra{I^-}\right)
    -\frac{1}{d(d^2-1)}\left(\ket{I^+}_{j\,j}\bra{I^-}
    +\ket{I^-}_{j\,j}\bra{I^+}\right)\right]\ket{0000}_j\\
    &=\bigotimes_{j=1}^N\left[\frac{\ket{I^+}_j+\ket{I^-}_j}{d(d+1)}\right].
  \end{aligned}
\end{equation}
The normalisation of the state gives
\begin{equation}
\braket{I^+|\mathbb{E}[\rho_0\otimes \rho_0]} = 
  \prod_{j=1}^N\left[\frac{{}_j\braket{I^+|I^+}_j + {}_j\braket{I^+|I^-}_j}{d(d+1)}\right] = 1,
\end{equation}
and therefore the initial conditions for the systems immediately follow from the definition of $\ket{G_{p,q,r,s}}$:
\begin{equation}\label{eq:init-cond-typical-sm}
  \braket{G_{p,q,r,s}|\mathbb{E}[\rho_0\otimes \rho_0]} 
  = \left(\frac{2d}{d(d+1)}\right)^{r+s} = \left(\frac{2}{3}\right)^{r+s}.
\end{equation}

\subsubsection{Steady-state solution}
Let us now derive the steady-state solution to the system \eqref{eq:ODE_system_RPP_q=2}, that is, the solution to:
\begin{equation}
  \frac{d G_{p,q,r,s}(t)}{dt} = 0.
\end{equation}
By inspecting the right-hand side of \eqref{eq:ODE_system_RPP_q=2}, one observes that the system admits the five linearly independent solutions:
\begin{equation}
\{ 2^p,\, 2^q,\, 2^r,\, 2^{-r},\, 1\}.
\end{equation}
The above functions span the space of steady-state solutions, since when the system \eqref{eq:ODE_system_RPP_q=2} is written in matrix form, $d\underline{{G}}(t)/dt= A\cdot \underline{G}$, the matrix $A$ has a five-dimensional kernel, therefore any steady-state solution can be expressed as a linear combination of the above stationary solutions,
\begin{equation}\label{eq:generalGinfty_sm}
G^{(\infty)}_{p,q,r,s} = \alpha 2^p + \beta 2^q + \gamma 2^r + \delta 2^{-s} + \epsilon,
\end{equation}
where $\{\alpha,\beta,\gamma,\delta,\epsilon\}$ are (so far) free parameters. To determine them, we recall that there must also exist $5$ integrals of motion (spanning the kernel of the \emph{transpose} of the matrix $A$), which can be understood as linear combinations of functions $G_{p,q,r,s}(t)$ that stay constant for any initial state,
\begin{equation}
  \mathcal{F}^{(j)}(t) = 
  \smashoperator{\sum_{\substack{p,q,r,s \\ p+q+r+s = N}}}
  \alpha^{(j)}_{p,q,r,s} G_{p,q,r,s}(t),\qquad 
  \frac{d\mathcal{F}^{(j)}(t)}{dt} = 0, \, \text{for }j=1,\ldots,5.
\end{equation}
In particular, these are given as
\begin{equation}
  \begin{gathered}
    \label{eq:integrals_motion_sm}
    \mathcal{F}^{(1)}(t) = G_{N,0,0,0}(t),\qquad
    \mathcal{F}^{(2)}(t) = G_{0,N,0,0}(t),\qquad
    \mathcal{F}^{(3)}(t) = G_{0,0,N,0}(t),\qquad
    \mathcal{F}^{(4)}(t) = G_{0,0,0,N}(t),\\
    \mathcal{F}^{(5)}(t) = 
    \smashoperator{\sum_{\substack{p,q,r,s \\ p+q+r+s = N}}}
    (-2)^s \frac{(p+q+r+s)!}{p! q! r! s!} G_{p,q,r,s}(t).
  \end{gathered}
\end{equation}
The first four of the above are obviously integrals of motion, whereas it is not immediate a priori that $d\mathcal{F}^{(5)}(t)/dt = 0$; however, this can be directly checked from~\eqref{eq:ODE_system_RPP_q=2}.

Using now the fact that the value of Eq.~\eqref{eq:integrals_motion_sm} is the same in the initial and stationary state, and combining it with the initial condition~\eqref{eq:init-cond-typical-sm}, we can determine the coefficients $\{\alpha,\beta,\gamma,\delta,\epsilon\}$ in Eq.~\eqref{eq:generalGinfty_sm} to finally yield 
\begin{equation}
\label{eq:steady-state-sol-typical-sm}
G^{(\infty)}_{p,q,r,s} = \frac{[1-2\left(\frac{2}{3}\right)^N+3^{-N}][2^{-(N-p)}+2^{-(N-q)}]+[\left(\frac{2}{3}\right)^N-2^{-(N-1)}+3^{-N}][1-2^{-(N-r)}+2^{-s}]}{[1-2^{-(N-1)}](1-2^{-N})}.
\end{equation}
The expression of the steady-state averaged purity is therefore given as
\begin{equation}
\label{eq:steady-state-purity-typical-sm}
\mathcal{P}_n^{(\infty)}= G^{(\infty)}_{n,N-n,0,0}= 
  \frac{[1-2\left(\frac{2}{3}\right)^N+3^{-N}][2^{-(N-n)}+2^{-n}]+[\left(\frac{2}{3}\right)^N-2^{-(N-1)}+3^{-N}][2-2^{-N}]}{[1-2^{-(N-1)}][1-2^{-N}]}.
\end{equation}

\subsubsection{Random Permutation Circuit with Arbitrary Initial States}
One can repeat an analogous analysis also in the case of generic initial states, but now the number of invariant states increases, and apart from $\{\ket{I^{0}},\ket{I^{x}},\ket{I^{+}},\ket{I^{-}}\}$ we need to introduce $11$ additional states $\{\ket{I^{j}}\}$ labelled with $j=1,\ldots,11$~\cite{bertini2025quantum},
\begin{equation}
\label{eq:15_states}
  \begin{gathered}
      \ket{I^{1}}=\sum_{a,b=0}^{q-1}\ket{baaa},\qquad
      \ket{I^{2}}=\sum_{a,b=0}^{q-1}\ket{abaa},\qquad
      \ket{I^{3}}=\sum_{a,b=0}^{q-1}\ket{aaba},\qquad
      \ket{I^{4}}=\sum_{a,b=0}^{q-1}\ket{aaab},\\
    \begin{aligned}
      \ket{I^{5}}&=\sum_{a,b,c=0}^{q-1}\ket{bcaa},&\qquad
      \ket{I^{6}}&=\sum_{a,b,c=0}^{q-1}\ket{baca},&\qquad
      \ket{I^{7}}&=\sum_{a,b,c=0}^{q-1}\ket{baac},\\
      \ket{I^{8}}&=\sum_{a,b,c=0}^{q-1}\ket{abca},&
      \ket{I^{9}}&=\sum_{a,b,c=0}^{q-1}\ket{abac},&
      \ket{I^{10}}&=\sum_{a,b,c=0}^{q-1}\ket{aabc},
    \end{aligned}\\
    \ket{I^{11}}=\sum_{a,b,c,d=0}^{q-1}\ket{abcd}.
  \end{gathered}
\end{equation}
Note that for conciseness we will interchangeably use $\ket{I^{12}}=\ket{I^0}$, $\ket{I^{13}}=\ket{I^x}$, $\ket{I^{14}}=\ket{I^-}$, $\ket{I^{15}}=\ket{I^+}$.
Using these one-site states we introduce a $N$-site basis of symmetric states 
\begin{equation}
  \ket{G_{\underline{n}}} := \frac{1}{N!}\sum_{\pi\in S_N}
  \pi
    \bigotimes_{j=1}^{15} \ket{I^{j}}^{\otimes n_j},\qquad
  \underline{n}=(n_1,n_2,\ldots,n_{15}),\quad n_1+n_2+\cdots + n_{15}=15,\quad n_j\ge 0,
\end{equation}
and the overlap functions
\begin{equation}
  G_{\underline{n}}(t) := \braket{G_{\underline{n}}|\mathbb{E}[\rho(t)\otimes \rho(t)]}.
\end{equation}
They satisfy a system of equations similar to~\eqref{eq:ODE_system_RPP_q=2},
\begin{equation}\label{eq:ODE_system_RP_general}
  \frac{d G_{\underline{n}}(t)}{dt} = \frac{2\lambda}{L-1}
  \sum_{j=1}^{15}\sum_{k=j+1}^{15}
  n_j n_k \sum_{i=1}^{15}
  \alpha_{j,k}^{i}\left(G_{\underline{n}_{j,k}^{i}}-G_{\underline{n}}\right),
\end{equation}
where we introduced $\underline{n}_{j,k}^{i}$ to mean
\begin{equation}
  \underline{n}_{j,k}^{i}=
  \begin{cases}
    (n_1,n_2,\ldots,n_j-1,\ldots,n_k-1,\ldots,n_i+2,\ldots,n_{15}),\quad& i\neq j,k,\\
    (n_1,n_2,\ldots,n_j+1,\ldots,n_k-1,\ldots,n_{15}),&i= j,\\
    (n_1,n_2,\ldots,n_j-1,\ldots,n_k+1,\ldots,n_{15}),&i=k,
  \end{cases}
\end{equation}
and we use the convention $G_{\underline{n}}=0$ if $n_j^{\prime}<0$ or $n_j^{\prime}>N$ for some $j^{\prime}$. The coefficients $\alpha_{j,k}^i$ are given as~\cite{bertini2025quantum},
\begin{equation}
  \alpha_{j,k}^i = \sum_{l=1}^{15} W_{i,l} 
  \braket{I^{l}|I^j} \braket{I^{l}|I^k},\qquad [W^{-1}]_{i,j}=\braket{I^{i}|I^{j}}^2.
\end{equation}

Even though compact, Eq.~\eqref{eq:ODE_system_RP_general} is not easy to study numerically due to a very high number of relevant states, however we can make some progress to study the stationary state. Let us focus on the $q=2$ case, as for $q\ge3$ the stationary states in case of two-site coupling coincides with the stationary state under global dynamics~\cite{gowers1996almost,chen2024incompressibility,gay2025pseudorandomness}.

To characterise the stationary state of Eq.~\eqref{eq:ODE_system_RP_general} we need to find all the integrals of motion, that is linear combinations of $G_{\underline{n}}$ which don't change for \emph{any} initial state, 
\begin{equation}
  \mathcal{F}=\sum_{\underline{n}} d_{\underline{n}} G_{\underline{n}},\qquad \frac{d\mathcal{F}(t)}{dt}=0.
\end{equation}
We can very quickly find $15$ of them by observing that the r.h.s.\ of~\eqref{eq:ODE_system_RP_general} has no terms with $j=k$, which gives
\begin{equation}
  \mathcal{F}^{(k)}(t)=G_{\underline{n}^{(k)}}(t),\qquad
  \underline{n}^{(k)}=(\underbrace{0,0,\ldots,0}_{k-1},N,\underbrace{0,0,\ldots,0}_{15-k}),
  \qquad k=1,\ldots 15.
\end{equation}
This is a consequence of a stronger property of the time-evolution, namely that applying any two-site gate on a state $\bra{I^{k}}^{\otimes N}$ from the right leaves the state invariant,
\begin{equation}
  \bra{I^{k}}^{\otimes N} \mathcal{U}_{i,j}= \bra{I^{k}}^{\otimes N}.
\end{equation}
Similarly, we observe that the following holds for all $\underline{n}$ and $j$, $k$
\begin{equation}
   \left(-2\bra{I^{0}} +\bra{I^{x}} +\bra{I^{+}} +\bra{I^{-}}\right)^{\otimes N}
   \left(\mathcal{U}_{j,k}-1\right)\ket{G_{\underline{n}}}=0,
\end{equation}
which gives the 16-th conservation law,
\begin{equation}
  \mathcal{F}^{(16)}(t)=\sum_{\underline{n}}
  \left(\prod_{j=1}^{11}\delta_{n_j,0}\right)
  \frac{(n_{0}+n_{x}+n_{+}+n_{-})!}{n_0!n_x!n_+!n_-!}
  (-2)^{n_0}
  G_{\underline{n}}(t).
\end{equation}
Note that $\mathcal{F}^{(16)}$ is analogous to $\mathcal{F}^{(5)}$ from~\eqref{eq:integrals_motion_sm}. While unable to prove that these are all the conservation laws for any system size $N$, analysis of the r.h.s.\ of Eq.~\eqref{eq:ODE_system_RP_general} for small $N$ suggests the kernel of the map is $16$-dimensional.

Similarly we find $16$ independent stationary states, $\underline{G}^{\infty}=\{G_{\underline{n}}^{\infty}\}_{\underline{n}}$, given by
\begin{equation}
  \begin{aligned}
    G_{\underline{n}}^{\infty,1}&=2^{n_1+n_5+n_6+n_7+n_{11}},&\qquad
    G_{\underline{n}}^{\infty,2}&=2^{n_2+n_5+n_8+n_9+n_{11}},\\
    G_{\underline{n}}^{\infty,3}&=2^{n_3+n_6+n_8+n_{10}+n_{11}},&\qquad
    G_{\underline{n}}^{\infty,4}&=2^{n_4+n_7+n_9+n_{10}+n_{11}},\\
    G_{\underline{n}}^{\infty,5}&=2^{-n_3-n_4+n_5+n_{11}-n_{12}-n_{13}-n_{15}},&\qquad
    G_{\underline{n}}^{\infty,6}&=2^{-n_2-n_4+n_6+n_{11}-n_{12}-n_{14}-n_{15}},\\
    G_{\underline{n}}^{\infty,7}&=2^{-n_2-n_3+n_7+n_{11}-n_{12}-n_{13}-n_{14}},&\qquad
    G_{\underline{n}}^{\infty,8}&=2^{-n_1-n_4+n_8+n_{11}-n_{12}-n_{13}-n_{14}},\\
    G_{\underline{n}}^{\infty,9}&=2^{-n_1-n_3+n_9+n_{11}-n_{12}-n_{14}-n_{15}},&\qquad
    G_{\underline{n}}^{\infty,10}&=2^{-n_1-n_2+n_{10}+n_{11}-n_{12}-n_{13}-n_{15}},\\
    G_{\underline{n}}^{\infty,11}&=2^{n_5+n_6+n_7+n_8+n_9+n_{10}+2n_{11}-n_{12}},&\qquad
    G_{\underline{n}}^{\infty,12}&=1,\\
    G_{\underline{n}}^{\infty,13}&=2^{n_6+n_9+n_{11}+n_{13}},&\qquad
    G_{\underline{n}}^{\infty,14}&=2^{n_5+n_10+n_{11}+n_{14}},\\
    G_{\underline{n}}^{\infty,15}&=2^{n_7+n_8+n_{11}+n_{15}},&\qquad
    G_{\underline{n}}^{\infty,16}&=2^{-n_1-n_2-n_3-n_4+n_{11}-n_{12}}.
  \end{aligned}
\end{equation}
Now we finally have everything needed to get a prediction for the stationary state under $2$-qubit dynamics,
\begin{equation}
  G_{\underline{n}}^{\infty}=\sum_{j=1}^{16} c_j G_{\underline{n}}^{\infty,j},
\end{equation}
with the coefficients $c_j$ determined by requiring $\mathcal{F}^{(j),\infty}=\mathcal{F}^{(j)}(0)$.

In the case of a random product initial state, the expression can be checked to reduce to the previous form. Assuming now that the initial state is not random, the stationary density depends on $4$  real initial-state dependent real parameters, and three phases, 
\begin{equation}\label{eq:relevant_parameters_sm}
  \begin{gathered}
    z(\ket{\psi}) e^{i \theta_1(\ket{\psi})} 
    = \frac{1}{\sqrt{d}} \sum_{s} \braket{s|\psi},\qquad
    y(\ket{\psi}) e^{i \theta_2(\ket{\psi})} = \sum_{s} \braket{s|\psi}^2,\\
    w(\ket{\psi}) e^{i \theta_3(\ket{\psi})} = \sum_{s} 
    |\braket{s|\psi}|^2 \braket{s|\psi},\qquad
    0\le z(\ket{\psi}),y(\ket{\psi}),w(\ket{\psi}) \le 1,\\
    Q(\ket{\psi})=\bra{\mathcal{F}^{(16)}}\left(\ket{\psi}^{\ast}\otimes \ket{\psi}\otimes
    \ket{\psi}^{\ast}\otimes \ket{\psi}\right),\qquad
    Q\ge 0,
  \end{gathered}
\end{equation}
and the stationary purity for an arbitrary non-random initial state is given by 
\begin{equation}
  \begin{aligned}
    P_{x}&=
    \frac{1}{(1-\frac{2}{d})(1-\frac{1}{d})(1-\frac{6}{d}+\frac{2}{d^2})}
    \Big\{
      \left(d^{-\frac{n}{N}}+d^{-\frac{N-n}{N}}\right)
      \left(1-\frac{4}{d}\right)\left(1-\frac{2}{d}-\frac{2}{d^2}\right)
      -\frac{4}{d}\left(1-\frac{11}{2 d}+\frac{8}{d^2} +\frac{1}{d^3}\right)\\
      &+\left(1-d^{-\frac{n}{N}}\right)\left(1-d^{-\frac{N-n}{n}}\right)
      \Big[\frac{Q}{d}\left(1-\frac{4}{d}\right)\left(1-\frac{2}{d}\right)
      I_2\left(1-\frac{6}{d}+\frac{14}{d^2}\right)
      - \frac{y^2}{d}\left(1-\frac{4}{d}+\frac{6}{d^2}\right)\\
      &+\frac{4 z^2}{d}\left(1-\frac{3z^2}{2}+\frac{2}{d}\right)
      +\frac{2yz^2}{d}\left(1+\frac{2}{d}\right)\cos(2\vartheta_1-\vartheta_2)
      -\frac{12}{d^{\frac{3}{2}}} w z \cos(\vartheta_1-\vartheta_3)
      \Big]
      \Big\},
  \end{aligned}
\end{equation}
where we have used the short-hand notation $d=2^N$. In the limit of large sizes the leading order contribution becomes for $x=n/N$
\begin{equation}
  P_{x} = d^{-x} + d^{-(1-x)} + (1-d^{-x}-d^{-(1-x)}) \left(I_2 + \frac{Q}{d}\right)
  + \mathcal{O}{d^{-1}}.
\end{equation}
For appropriate values of $Q$, this does not necessarily coincide with the global-permutation result (cf.\ \eqref{eq:globalPermutationPageNonRandom_sm}). For example, for homogeneous non-random product initial states one gets
\begin{equation}
  \frac{Q}{d}-z^4\ge 0,\qquad \frac{Q}{d}-z^4 = \mathcal{O}(1),
\end{equation}
and therefore we do not necessarily saturate the bound~\eqref{eq:final_bound} even in the limit of large system sizes.  

\section{Global permutation ensembles}
In this section, we compute the average R\'enyi-2 entropy in the ensemble $\mathcal{E}_\text{GPP}$ (global phase transformations followed by global permutations) and --- for a locally scrambled state --- in the ensemble $\mathcal{E}_\text{GP}$ (global permutations). Namely, we compute the average R\'enyi-2 entropy of $U^\text{RPP}\ket{\psi_0}$ and $U^\text{RP}\ket{\psi_0}$, where $U^\text{RPP}$, $U^\text{RP}$ are $d \times d$ matrices drawn from the two ensembles. We then show that for the ensemble $\mathcal{E}_\text{GPP}$ the result coincides with that obtained in the ensemble of infinite-depth circuits $\mathcal{E}_\text{PPC}(D\to \infty)$ at any system size, while for the ensemble $\mathcal{E}_\text{GP}$ the average R\'enyi-2 entropy coincides with that in $\mathcal{E}_\text{PC}(D\to \infty)$ only as $N\to \infty$.

\subsection{Page curves}
To begin with, we consider a tripartite Hilbert space $\mathcal{H}=\mathcal{H}_-\otimes \mathcal{H}_+\otimes \mathcal{H}_0$ as in \eqref{eq:tripartite_Hilbert_space}, associated to three regions $A_-$, $A_+$, $A_0$ with $|A_-|=n_-$, $|A_+|=n_+$, $|A_0|=n_0$. The dimension of the Hilbert space is: 
\begin{equation}
d=d_-\,d_+\,d_0 =q^{n_-}q^{n_+}q^{n_0}=q^N.
\end{equation}
For $\alpha=-,+,0$, we let $\ket{I^\alpha}=\otimes_{j\in A_\alpha}\ket{I^\alpha}_j$ and we consider the overlaps $\braket{{I^\alpha}|\rho_0\otimes\rho_0}$ for a pure state $\rho_0 =\ket{\psi_0}\bra{\psi_0}$. These can be obtained from the expression~\eqref{eq:G_functions_IPR_representation},
\begin{equation}
  \begin{aligned}
    G_{N,0,0}(\ket{\psi_0})&=\braket{{I^-}|\rho_0 \otimes\rho_0}
    =\text{Tr}_{\mathcal{H}}(\rho_0)^2 = 1, \\
    G_{0,N,0}(\ket{\psi_0})&=\braket{{I^+}|\rho_0 \otimes\rho_0}
    =(\text{Tr}_{\mathcal{H}}\,\rho_0)^2 = 1, \\
    G_{0,0,N}(\ket{\psi_0})&=\braket{{I^0}|\rho_0 \otimes\rho_0}
    =\sum_{\{x_i\}} |\braket{\{x_i\}|\psi_0}|^4 = I_2(\ket{\psi_0}).
  \end{aligned}
\end{equation}
The above expressions are valid for a generic initial state $\ket{\psi_0}$. From now on, we specialize to the case of permutation-invariant states, namely:
\begin{enumerate}
    \item Product states of the form \eqref{eq:homogeneous_product_state}, for which
    $I_2(\ket{\psi_0})= \left(\sum_{a=0}^{q-1}|\lambda_a|^4\right)^N$;
    \item The Dicke states $\ket{D}$, eq. \eqref{eq:Dicke_state_def}, for which $I_2(\ket{D})= \sum_{m=1}^N|c_m|^4\binom{N}{m}^{-1};$
    \item The GHZ state $\ket{\text{GHZ}}$, for which $I_2(\ket{\text{GHZ}})=q^{-1}$.
\end{enumerate}
The states above are chosen as they provide three different scaling laws for the Inverse Participation Ratio, with a localization which is exponential, polynomial, and constant in the system size, respectively. 

Let us consider the ensemble $\mathcal{E}_\text{GPP}$. The average purity of the state $U^\text{RPP}\ket{\psi_0}$ is obtained by computing the quantities:
\begin{equation}
  \mathcal{G}_{n_-,n_+,n_0}(\ket{\psi_0}) := \braket{G_{n_-,n_+,n_0}|\mathcal{U}^\text{RPP}|\rho_0\otimes\rho_0},
\end{equation}
where
\begin{equation}
  \begin{aligned}
    \mathcal{U}^\text{RPP} :&= \mathbb{E}\left[(U^\text{RPP})^*\otimes U^\text{RPP} 
    \otimes (U^\text{RPP})^*\otimes U^\text{RPP}\right] \\
    &= \frac{1}{d(d-1)}\left[\ket{I^-}\!\bra{I^-} + \ket{I^+}\!\bra{I^+} 
    + (d+1)\ket{I^0}\!\bra{I^0} - \ket{I^-}\!\bra{I^0}
    -\ket{I^0}\!\bra{I^-}-\ket{I^+}\!\bra{I^0}-\ket{I^0}\!\bra{I^+}\right],
  \end{aligned}
\end{equation}
is the ensemble average.

From the expression \eqref{eq:G_state_RPP_permutation_explicit}, we obtain
\begin{equation} \label{eq:global_G-functions_RPP}
  \begin{aligned}
    &\mathcal{G}_{n_-,n_+,n_0}(\ket{\psi_0}) 
    = \frac{1}{N!}\sum_{\pi \in S_N}\,\bigotimes_{j \in A_\pi^-}\bra{I^-}_j
    \bigotimes_{j \in A_\pi^+}\bra{I^+}_j\bigotimes_{j \in A_\pi^0}\bra{I^0}_j\, \\
    &\frac{1}{d(d-1)}\left[\ket{I^-}\!\bra{I^-} 
    + \ket{I^+}\!\bra{I^+} + (d+1)\ket{I^0}\!\bra{I^0} 
    - \ket{I^-}\!\bra{I^0}-\ket{I^0}\!\bra{I^-}-\ket{I^+}\!\bra{I^0}
    -\ket{I^0}\!\bra{I^+}\right]\ket{\rho_0 \otimes \rho_0} \\
    =&\frac{1}{d-1}
    \left[(d_-+d_{+})(1-I_2(\ket{\psi_0}))+ (d+1)I_2(\ket{\psi_0})-2\right].
  \end{aligned}
\end{equation}
Notice that $\mathcal{G}_{n,N-n_0 - n, n_0}=\mathcal{G}_{N-n_0 - n,n, n_0},$ $\forall n_0 = 0,\dots,N-n$, which entails the known symmetry of the purity ($n_0=0$) $\mathcal{P}_n = \mathcal{P}_{N-n}$.

We note that the expression~\eqref{eq:global_G-functions_RPP} is surprisingly general: for \emph{every} permutation-invariant initial state $\ket{\psi_0}$ the function $\mathcal{G}_{n_-,n_+,n_0}(\ket{\psi_0})$ depends only on the (sub)system sizes $N$, $n$, on the Hilbert-space dimension $d$, and on $I_2(\ket{\psi_0})$, the latter being the only state-dependent quantity.
Letting $x=\frac{n_-}{N}$, $1-x = \frac{n_+}{N}$, at large $N$ the average purity at leading order in $N$ can be read out from the above expression:
\begin{equation}
\mathcal{P}^\text{GPP}_x \simeq q^{-x N} + q^{(x-1)N} + I_2(\ket{\psi})[1-q^{-x N} - q^{(x-1)N}], 
\end{equation}
which implies
\begin{equation}
-\frac{1}{N} \log \mathcal{P}_x^\text{GPP} = \log (q) \,\text{min}\left\{x,1-x,-\frac{\log I_2(\ket{\psi})}{N \log q}\right\} + O\left(\frac{1}{N}\right)\,.
\end{equation}
From the above expression we can define a condition of \textit{high localization}: if $\exists \,x < 1/2$ such that the inequality
\begin{equation}
-\frac{\log I_2(\ket{\psi})}{N \log q} < x\,,
\end{equation}
is verified, then maximal scrambling is prevented in the state $\ket{\psi}$, meaning that the Page curve does not reproduce that of the Haar random ensemble at large system size. This condition is always satisfied for the product state with non-maximal tilting ($\theta \ne \pi/4$ for $q=2$), for the GHZ state and for the Dicke states.

\subsection{Proof of equivalence at late times}
Here, we show that the functions $\mathcal{G}_{n_-,n_+,n_0}[\ket{\psi_0}]$  obtained in \eqref{eq:global_G-functions_RPP} solve the steady-state form of the system \eqref{eq:purity_ODE_system}, which is obtained from the latter by removing the time derivative term. We do so by first observing that $d$ and $I_2(\ket{\psi_0})$ is invariant under the changes of $n_{-}$, $n_{+}$, $n_{0}$ that preserve their sum. Therefore we have
\begin{equation}
  \begin{aligned}
    &\mathcal{G}_{n_--1,n_++1,n_0} + \mathcal{G}_{n_-+1,n_+-1,n_0} 
    + (q-1)\mathcal{G}_{n_--1,n_+-1,n_0+2} \\
    =&\frac{q^{-1}\,d_-(1-\text{I}_2)+q\,d_+(1-\text{I}_2)+[(d+1)\text{I}_2 -2]}{d-1} 
    + \frac{q\,d_-(1-\text{I}_2)+q^{-1}\,d_+(1-\text{I}_2)+[(d+1)\text{I}_2 -2]}{d-1} \\
    +&(q-1)\frac{q^{-1}\,d_-(1-\text{I}_2)+q^{-1}\,d_+(1-\text{I}_2)
    +[(d+1)\text{I}_2 -2]}{d-1} \\
    =&(q+1)\mathcal{G}_{n_-,n_+,n_0}\,,
  \end{aligned}
\end{equation}
and analogously
\begin{equation*}
\mathcal{G}_{n_-+1,n_+,n_0-1} + q\,\mathcal{G}_{n_--1,n_+,n_0+1} = \mathcal{G}_{n_-,n_++1,n_0-1} + q\,\mathcal{G}_{n_-,n_+-1,n_0+1} = (q+1)\mathcal{G}_{n_-,n_+,n_0}.
\end{equation*}
A simple inspection of equation \eqref{eq:purity_ODE_system} then reveals that
\begin{equation}
G_{n_-,n_+,n_0}(t\to \infty) = \mathcal{G}_{n_-,n_+,n_0}(\ket{\psi_0})
\end{equation}
is a steady state solution for every choice of the permutation-invariant initial state $\ket{\psi_0}$.

\subsection{Page curves for the $\mathcal{E}_{GP}$}
Finally, let us consider the averaged purity in the ensemble $\mathcal{E}_{GP}$, i.e.\ the averaged purity after acting with an all-to-all permutation gate on an initial state. As before, we use $d$ to denote the full Hilbert-space dimension $d=q^N$, and we are interested in a bipartition of the system into subsystems with $n$ and $N-n$ sites. Similarly to before, the central object is the averaged $4$-replica gate $\mathcal{U}^{\mathrm{RP}}$, defined as
\begin{equation}
  \mathcal{U}^{\mathrm{RP}} := 
  \mathbb{E}\left[(U^{\mathrm{RP}})^\ast \otimes U^{\mathrm{RP}} \otimes 
    (U^{\mathrm{RP}})^\ast \otimes U^{\mathrm{RP}}\right].
\end{equation}
The matrix $\mathcal{U}^{\mathrm{RP}}$ is again a projector to the set of invariant states, but the number of invariant states is now $15$. Apart from $\ket{I^{\alpha}}$ for $\alpha\in\{0,+,-,x\}$, we also have $11$ additional states $\ket{I^{j}}$, $j=1,\ldots,11$, as introduced in Eq.~\eqref{eq:15_states}.  Using now the fact that $\mathcal{U}^{\mathrm{RP}}$ is a projector to the space spanned by $\ket{I^{\alpha}}^{\otimes N}$ for $\alpha\in\{1,2,\ldots,11,0,x,+,-\}$, we can express the averaged purity as
\begin{equation}\label{eq:purityGPdefinition}
  \mathcal{P}_n^{\mathrm{GP}}=\smashoperator{\sum_{\alpha,\beta\in\{1,2,\ldots,11,0,x,+,-\}}}
  W_{\alpha,\beta}(N)
  \braket{I^{+}|I^{\alpha}}^n
  \braket{I^{-}|I^{\alpha}}^{N-n}
  \left(\bra{I^{\beta}}^{\otimes N}\right)\ket{\rho_0 \otimes \rho_0},
\end{equation}
where $W(N)$ is the inverse of the matrix of overlaps
\begin{equation}
  \left[W^{-1}(N)\right]_{\alpha,\beta}=\braket{I^{\alpha}|I^{\beta}}^N.
\end{equation}
All the overlaps between states $\ket{I^{\alpha}}$ and the initial state are given in terms of the amplitudes and phases given in Eq.~\eqref{eq:relevant_parameters_sm}, as well as the inverse participation ratio $I_2(\ket{\psi})$ (cf.\ \eqref{eq:inverse_participation}). Explicitly, the overlaps are given as
\begin{equation}
  \mkern-30mu
  \begin{aligned}
    \bra{I^{0}}^{\otimes N}\ket{\rho_0\otimes\rho_0}&=I_2(\ket{\psi}),\qquad
  \bra{I^{x}}^{\otimes N}\ket{\rho_0\otimes\rho_0}=y(\ket{\psi})^2,\qquad
  \bra{I^{+}}^{\otimes N}\ket{\rho_0\otimes\rho_0}=
  \bra{I^{-}}^{\otimes N}\ket{\rho_0\otimes\rho_0}=1,\\
    \bra{I^{1}}^{\otimes N}\ket{\rho_0\otimes\rho_0}&=
    \bra{I^{2}}^{\otimes N}\ket{\rho_0\otimes\rho_0}^{\ast}=
    \bra{I^{3}}^{\otimes N}\ket{\rho_0\otimes\rho_0}=
    \bra{I^{4}}^{\otimes N}\ket{\rho_0\otimes\rho_0}^\ast
    =\sqrt{d}z(\ket{\psi}) w(\ket{\psi}) e^{i [\theta_3(\ket{\psi})-\theta_1(\ket{\psi})]},\\
    \bra{I^{5}}^{\otimes N}\ket{\rho_0\otimes\rho_0}&=
    \bra{I^{7}}^{\otimes N}\ket{\rho_0\otimes\rho_0}=
    \bra{I^{8}}^{\otimes N}\ket{\rho_0\otimes\rho_0}=
    \bra{I^{10}}^{\otimes N}\ket{\rho_0\otimes\rho_0}=d z(\ket{\psi})^2,\\
    \bra{I^{6}}^{\otimes N}\ket{\rho_0\otimes\rho_0}&=
    \bra{I^{9}}^{\otimes N}\ket{\rho_0\otimes\rho_0}^{\ast}=d z(\ket{\psi})^2 y(\ket{\psi})
    e^{i[\vartheta_2(\ket{\psi})-2\vartheta_1(\ket{\psi})]},\\
    \bra{I^{11}}^{\otimes N}\ket{\rho_0\otimes\rho_0}&=d^2 z(\ket{\psi})^4.
  \end{aligned}
  \mkern-30mu
\end{equation}
Inserting this into the equation for averaged purity gives the following general expression,
\begin{equation}
\begin{aligned}
  \mathcal{P}_n&=\frac{\left(1-\frac{4}{d}\right)\left(q^{-(N-n)}+q^{-n}\right)-\frac{2}{d}\left(1-\frac{5}{d}\right)}{\left(1-\frac{3}{d}\right)\left(1-\frac{2}{d}\right)}
  +\frac{\left(1-q^{-(N-n)}\right) \left(1-q^{-n}\right)}{\left(1-\frac{1}{d}\right)
    \left(1-\frac{2}{d}\right)\left(1-\frac{3}{d}\right)}
    \Bigg[
    \left(1-\frac{5}{d}\right) I_2(\ket{\psi})
    +\frac{1}{d^2} y(\ket{\psi})\\
    &+8 d^{-\frac{3}{2}} w(\ket{\psi}) z(\ket{\psi})
    \cos\left[\vartheta_1(\ket{\psi})-\vartheta_3(\ket{\psi})\right]
    -z(\ket{\psi})^2\frac{2}{d}\left(2+y(\ket{\psi})
    \cos\left[2\vartheta_1(\ket{\psi})-\vartheta_2(\ket{\psi})\right]\right)
    +z(\ket{\psi})^4\Bigg].
\end{aligned}
\end{equation}
Note that this expression is completely general and holds for any initial state. For convenience let us also explicitly consider an average over Haar-random initial product states, which is obtained from the general form~\eqref{eq:purityGPdefinition} upon replacing the initial-state overlap by an averaged initial state overlap,
\begin{equation}
  \left(\bra{I^{\beta}}^{\otimes N}\right)\ket{\rho_0\otimes\rho_0} \mapsto 
  \left(\bra{I^{\beta}}^{\otimes N}\right)\mathbb{E}[\ket{\rho_0\otimes\rho_0}],
\end{equation}
and these can be straightforwardly obtained by noting
\begin{equation}
  \mathbb{E}[\ket{\rho_0\otimes\rho_0}]
  =\bigotimes_{j=1}^{N} \frac{1}{q(q+1)}\left(\ket{I^{+}}+\ket{I^{-}}\right).
\end{equation}
Explicitly evaluating the overlaps and plugging it into Eq.~\eqref{eq:purityGPdefinition} gives (after some straightforward manipulations) the general form
\begin{equation}
  \mathbb{E}[\mathcal{P}_{n}]=
  \frac{(q^{-n}+q^{-(N-n)})(1-\frac{5}{d}+\frac{8}{d^2})+
  \left((1-q^{-n})(1-q^{-(N-n)})\mathbb{E}[I_2]-\frac{2}{d}\right)(1-\frac{4}{d}+\frac{7}{d^2})}
  {(1-\frac{1}{d})(1-\frac{2}{d})(1-\frac{3}{d})}, 
\end{equation}
where we have used $\mathbb{E}[I_2]$ to denote the value of $I_2$ averaged over the random product initial state,
\begin{equation}
  \mathbb{E}[I_2]=\left(\frac{2}{q+1}\right)^N.
\end{equation}
Note that specialising to $q=2$ this gives the expression in Eq.~\eqref{eq:purity_global_average}.

In the limit of large $N$ with $n/N=x$ fixed, the two expressions reduce to
\begin{equation}\label{eq:globalPermutationPageNonRandom_sm}
  \mathcal{P}_x = q^{- x N} +q^{-(1-x)N}
  +\left(1 - q^{- x N} - q^{-(1-x)N}\right) \left[I_{2}(\ket{\Psi})+z(\ket{\psi})^4\right]
  +\mathcal{O}(q^{-N}),
\end{equation}
and
\begin{equation}
  \mathbb{E}[\mathcal{P}_x]=
  q^{-x N}+q^{-(1-x)N}+\left(1-q^{-x N}-q^{-(1-x)N}\right)\mathbb{E}[I_2]
  +\mathcal{O}(q^{-N}).
\end{equation}

\section{Bosonic-space formulation}\label{sec:bosonic-sm}
An alternative way of solving the dynamics of the circuit is to map the problem to a bosonic Hilbert space by taking advantage of the (site) permutation-invariance of the dynamics and the initial state. Following Ref.~\cite{piroli2020random}, we define site permutation-invariant states of the form

\begin{equation} \label{eq:bosonic-basis-sm}
  \ket{n_-, n_+, n_x, n_0} := \frac{1}{\sqrt{N!\,n_-!\,n_+!\,n_x!\,n_0!}}
  \sum_{\pi\in \mathcal{S}_N}\pi \ket{J^-}^{\otimes n_-}\otimes 
  \ket{J^+}^{\otimes n_+}\otimes \ket{J^x}^{\otimes n_x}\otimes 
  \ket{J^0}^{\otimes n_0}
\end{equation}
for every $n_-, n_+, n_x, n_0 \geq 0$ such that they sum to $N$. In the above formula, $\mathcal{S}_N$ is the permutation group over $N$ elements and $\pi$ is the unitary operator corresponding to an element of $\mathcal{S}_N$. The states \eqref{eq:bosonic-basis-sm} are (site) permutation-invariant and form an orthonormal basis by construction. As a result, they can be rewritten in terms of bosonic creation and annihilation operators acting on a vacuum $\ket{\Omega}$
\begin{equation}
  \ket{n_{-},n_{+},n_{x}, n_{0}} := 
  \frac{1}{\sqrt{n_-!\,n_+!\,n_x!\,n_0!}}
  \Big(a_{-}^\dagger\Big)^{n_{-}}\Big(a_{+}^\dagger\Big)^{n_{+}}
  \Big(a_{x}^\dagger\Big)^{n_{x}}\Big(a_{0}^\dagger\Big)^{n_{0}}\ket{\Omega}\,,
\end{equation}
which satisfy the usual canonical commutation relations
\begin{equation}
  [a_\alpha, a_\beta^\dagger] = \delta_{\alpha,\beta}\,,\qquad  [a_\alpha, a_\beta] 
  =  [a_\alpha^\dagger, a_\beta^\dagger] = 0\,.
\end{equation}
The Lindbladian~\eqref{eq:Ujk-rpc-typical-sm} can be expressed in terms of these operators by using~\cite{piroli2020random}
\begin{equation}
  \sum_{j = 1}^{N}(\ket{x}\!\bra{y})_j = a_x^\dagger a_y
\end{equation}
and
\begin{equation}
    \sum_{j<k}\sum_{x, y, z, t}\Big[\Gamma^{x,y}\ket{xy}_{j,k}\Big]\Big[\Lambda^{z,t}{}_{j,k}\!\bra{zt}\Big] = \frac{1}{2}\sum_{x, y, z, t}\Gamma^{x,y}\Lambda^{z,t}\left[a_x^\dagger a_y^\dagger a_z a_t\right]
\end{equation}
for every pair of symmetric matrices $\Gamma, \Lambda$. Here, $x, y, z, t$ run over the different ``species'' of bosons, $-, +, x$ and $0$. Now the operators appearing in the above formula can be simply matched to those of Eq.~\eqref{eq:Ujk-rpc-typical-sm}, allowing us to express the averaged gate in terms of the global bosonic operators:
\begin{equation}
\begin{split}
    &\sum_{1\leq i<j\leq N}\tilde{\mathcal{U}}_{i,j} = \sum_{1\leq i<j\leq N}\Pi_i\Pi_j\, \mathcal{U}_{i,j}\Pi_i\Pi_j =\\
    &\frac{1}{2}n_0(n_0-1) + \frac{1}{6}\bigg[2n_-(n_--1)+2n_+(n_+-1)+2n_x(n_x-1)+4n_0n_- + 4n_0n_+  + 4n_0n_x\\
    & +2a_-^\dagger n_- a_0 + 2a_0^\dagger n_- a_-+ 2a_+^\dagger n_+ a_0 + 2a_0^\dagger n_+ a_+ + 2a_x^\dagger n_x a_0 + 2a_0^\dagger n_x a_x\bigg] \\
    &+\frac{1}{3}\bigg[n_-n_+ + n_- n_x + n_+ n_x +a_+^\dagger n_-  a_x + a_+^\dagger n_+  a_x + a_x^\dagger n_-  a_x + a_-^\dagger n_x  a_+ + a_x^\dagger n_+  a_- + a_+^\dagger n_x  a_-\bigg]\,,\label{eq:lind-rpc-typical-bosonic-sm}
\end{split}
\end{equation}
where we have introduced the ``particle number'' operators
\begin{equation}
  n_{-} = a_{-}^\dagger a_{-},\qquad
  n_{+} = a_{+}^\dagger a_{+},\qquad
  n_{x} = a_{x}^\dagger a_{x},\qquad
  n_{0} = a_{0}^\dagger a_{0}\,.
\end{equation}
The operator \eqref{eq:lind-rpc-typical-bosonic-sm} can be represented by a finite (sparse) matrix in the basis \eqref{eq:bosonic-basis-sm}, allowing for simple, numericaly exact simulation of the dynamics starting from typical initial states given by \eqref{eq:typical-init-state_sm}. These states are easily represented in the bosonic basis and take the form
\begin{equation}
    \ket{\mathbb{E}[\rho_0\otimes \rho_0]} 
    = \frac{1}{6^N}\sum_{n_+, n_-, n_0} 
    2^{N/2+n_0}\sqrt{\frac{N!}{n_-!n_+!n_0!}}\ket{n_{-},n_{+},0, n_{0}}\,,
\end{equation}
where the sum is over non-negative integers $n_-, n_+, n_0$ such that their sum is $N$. Similar computation yields the expression for the symmetrized boundary (swap) state
\begin{equation}
\begin{aligned}
  \ket{X_n} :=& \frac{1}{N!} \sum_{\pi \in \mathcal{S}_N} 
  \pi \ket{I^-}^{\otimes n} \otimes \ket{I^+}^{\otimes N-n}\\
       =& \sum_{n_{+}=0}^{N-n}\sum_{n_{-}=0}^{n} 2^{N/2}
       \frac{ n! (N-n)!}{\left(n-n_-\right)! \left(N-n -n_+\right)!}
       \sqrt{\frac{\left(N-n_- -n_+\right)!}{N! n_-! n_+!}} \ket{n_{-},n_{+},0,N-n_+-n_-}\,.
\end{aligned}
\end{equation}
One can easily check that $\braket{X_n|\rho_0\otimes\rho_0}=1$.

\section{Numerical results}

We now provide some details about the numerical implementation of the dynamics as well as some additional results.

Solving the dynamics of the Rényi entropies for finite system sizes amounts to solving the differential equations \eqref{eq:differential_system_main} together with appropriate initial conditons corresponding to typical initial states \eqref{eq:init-cond-typical-sm}. Eq. \eqref{eq:differential_system_main} corresponds to a set of coupled linear ordinary differential equations. As such, the system can be straightforwardly solved via well-known methods. Importantly, however, when the system size $N$ grows above a certain value (in our calculations, this was $N\gtrsim 32$) numerical instabilities appear and the results (unphysically) diverge.

A way around this is given by the bosonic formulation of the problem detailed in App. \ref{sec:bosonic-sm}. Mappig the dynamics to the bosonic Hilbert space allows for exact (numerical) vector representation of the system states and the Lindbladian encoding the dynamics. As the Lindbladian has only positive eigenvalues, unphysical divergences are not a problem and it is possible to solve much larger system sizes. In our calculations, we managed to go up to $N = 168$.
In this approach, the most computationally demanding part is the exponentialization of the Lindbladian matrix. One can, however, take advantage of the sparse nature of $\mathcal{L}$ and solve the differential equation corresponding to \eqref{eq:projected_dynamics} via simple numerical methods, such as Euler or Runge-Kutta type methods. For random permutation circuits starting from typical initial states, all of our results are obtained via forward Euler with 12 digit precision, as apparent from comparison with the 4th order Runge-Kutta solutions.

\begin{figure}[h]
    \centering
    \includegraphics[width=.8\linewidth]{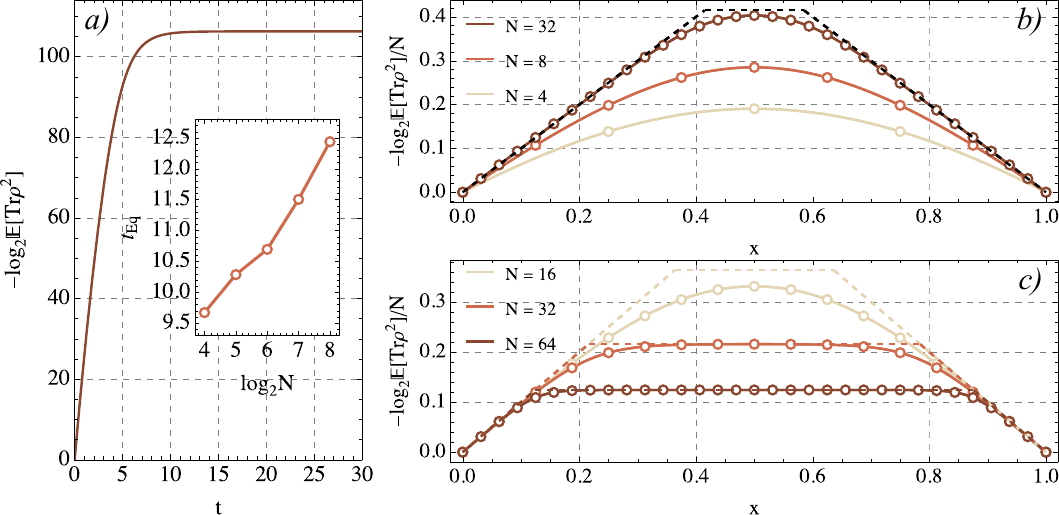}
    \caption{Rényi-2 entropy dynamics in random permutation circuits with additional random phases. \textit{a)} The time evolution of the averaged Rényi-2 entropy from the initial product state $\ket{\psi}=\ket{\phi(\theta)}^{\otimes N}$, with $\ket{\phi(\theta)}=\cos\theta|0\rangle+\sin(\theta)|1\rangle$ and $\theta = \pi/8$. The system size is set to $N = 256$. \textit{Inset:} the equilibration time as a function of the logarithm of the system size $N$. \textit{b)} Comparison of the Page curves obtained by averaging over the random circuit (open markers) and the random global (solid lines) ensemble starting from the product state corresponding to $\theta = \pi/8$. The dashed lines show the (tight) upper bounds. \textit{c)} Same as \textit{b)}, starting from the initial state $\ket{D} = (\ket{D_1}+\ket{D_2})/\sqrt{2}$. It is clear that the global and circuit results coincide for finite $N$, as opposed to the case with no random phases.}
    \label{fig:rppc-sm}
\end{figure}

For different initial states, such as product states $\ket{\psi}=\ket{\phi(\theta)}^{\otimes N}$ with $\ket{\phi(\theta)}=\cos\theta|0\rangle+\sin(\theta)|1\rangle$ and fixed $\theta$, the corresponding differential equations are generally very complicated and is restricted to using numerical exact diagonalisation (ED) techniques, which are limited to small system sizes $N\lesssim 10$. In particular, we simulate the random dynamics by sampling different random realizations of the circuits; as such, the finite sample size of the gates result in some numerical error. All numerical methods detailed in the paper were crosschecked by ED for the available system sizes, together with the analytical global ensemble results.

We end this section by presenting an example of data generated by studying the dynamics of random-permutation circuits incorporating additional random phases, as discussed in the main text. The random phases simplify the problem, and one can derive the differential equations \eqref{eq:purity_ODE_system}, independently of the initial state of the system. As a result, the Rényi entropy dynamics of the RPP circuit can be studied for both product and entangled initial states, see Fig. \ref{fig:rppc-sm}. Moreover, integrating the differential equations did not give rise to numerical instabilities for the system sizes studied, and all results presented here are obtained via solving \eqref{eq:purity_ODE_system} (up to $N = 256$). From our plots, we see the results anticipated in the main text. Notably, we see from Fig.~\ref{fig:rppc-sm} that the Page curves corresponding to the late-time circuit ensemble and the global ensemble coincide for all finite-$N$, and that they are both bounded by the PEs.

\end{document}